\documentclass[11pt, letterpaper]{article}

\usepackage[pdftex]{graphicx}
\usepackage[latin1]{inputenc}
\usepackage{amsmath}
\usepackage{amsfonts}
\usepackage{amssymb}
\usepackage{amsthm}
\usepackage{algorithm}
\usepackage{algorithmic}
\usepackage{xcolor}
\usepackage{url}
\usepackage{natbib}

\usepackage{JASA_manu}

\usepackage{tikz}
\def\check{\tikz\fill[scale=0.4](0,.35) -- (.25,0) -- (1,.7) -- (.25,.15) -- cycle;} 

\theoremstyle{plain}

\theoremstyle{definition}

\theoremstyle{remark}

\include{header}
\numberwithin{theorem}{section} 

\usepackage{authblk}
\date{}
\begin{document}
\title {Measuring Influence in Twitter Ecosystems\\using a Counting Proccess Modeling Framework}
\author[1]{Donggeng Xia\footnote{Authors contributed equally to this work}}
\author[2]{Shawn Mankad$^*$}
\author[1]{George Michailidis}
\affil[1]{Department of Statistics, University of Michigan}
\affil[2]{Department of Decision, Operations \& Information Technologies, University of Maryland}
\renewcommand\Authfont{\scshape\small}
\renewcommand\Affilfont{\itshape\small}
\setlength{\affilsep}{0em}

\maketitle


\begin{center}
\textbf{Abstract}
\end{center}
Data extracted from social media platforms, such as Twitter,  are both large in scale and complex in nature, since they contain both unstructured text, as well as structured data,
such as time stamps and interactions between users. A key question for such platforms is to determine influential users, in the sense that they generate interactions between members of
the platform. Common measures used both in the academic literature and by companies that provide analytics services are variants of the popular web-search PageRank algorithm 
applied to networks that capture connections between users. In this work, we develop a modeling framework using multivariate interacting counting processes to capture the detailed actions
that users undertake on such platforms, namely posting original content, reposting and/or mentioning other users' postings. Based on the proposed model, we also derive a novel influence measure.
We discuss estimation of the model parameters through maximum likelihood and establish their asymptotic properties. The proposed model and the accompanying influence measure  are illustrated on 
a data set  covering a five year period of the Twitter actions of the members of the US Senate, as well as mainstream news organizations and media personalities.



%


\newpage
\setcounter{page}{1}    
\pagenumbering{arabic}  



\section{Introduction}\label{intro}

Leading business and non-profit organizations are integrating growing
volumes of increasingly complex {\em structured} and {\em
  unstructured} data to create big data ecosystems for content
distribution, as well as to gain insights for decision making.  A recent,
substantial area of growth has been online review and social media
platforms, which have fundamentally altered the public discourse by
providing easy to use forums for the distribution and exchange of
news, ideas and opinions.  The focus in diverse areas, including
marketing, business analytics and social network analysis, is to
identify trends and extract patterns in the vast amount of data
produced by these platforms, so that more careful targeting of content
distribution, propagation of ideas, opinions and products, as well as
resource optimization is achieved.

One platform that has become of central importance to both business
and non-profit enterprises is Twitter. According to its second quarter
2014 financial results announcement, Twitter had more than half a
billion users in July 2014, out of which more than 271 million were
active ones \citep{twitterStatistics}. Although Twitter lags behind in
terms of active users to Facebook, it is nevertheless perceived by
most businesses and non-profit organizations as an integral part of
their digital presence \citep{bulearca2010twitter}.

The mechanics of Twitter are as follows: the basic communication unit
is the account.  The platform allows account users to post messages of
at most 140 characters, and thus has been described as the Short
Message Service (SMS) of the Internet.  As of mid-2014, over half a
billion messages were posted on a daily basis.  Further, Twitter
allows accounts to ``follow'' other accounts, which means the follower
receives notification whenever the followed account posts a new
message.  Thus, the follow-follower relations serve as a primary
channel for content to spread within the social networking platform.
Accounts tend to interact with each other over these channels in two
directed ways.  First, an account can {\em copy} or {\em rebroadcast}
another account's tweet, which is referred to as a
``retweeting''. Second, an account can {\em mention} another account
within a tweet by referring to their account name with the $@$ symbol
as a prefix. These two actions, retweeting and mentioning, are
directed responses from one account to another and thus, provide the
mechanisms for online conversation.

The mechanics of Twitter, together with the original messages
generated by users, give rise to rich Big Data.  Specifically, the
content of the message, together with easily searchable key terms or
topics that use the \# symbol as a prefix, constitute a large
corpus of unstructured text. The hashtag function enables searches to
identify emerging themes and topics of discussion. In 2014, more than
2.1 billion search queries were generated \citep{twitterStatistics}.
Further, the following built-in capability, creates a network for {\em
  potential information flow and dissemination}, while the retweeting
and mentioning actions create subnetworks of {\em actual interactions}
between user accounts.

A key problem in all social networking platforms is that of
identifying {\em user influence}, since such users are capable of
driving action (e.g. steer discussions to particular themes and
topics) or provoking interactions amongst other users. Further, users
exhibiting high influence are also potentially more valuable to
businesses \citep{trusov2010determining}. The ranking of Twitter users
based on their influence constitutes both an active research topic and
a business opportunity, as manifested by services such as Klout
\citep{Klout} and PeerIndex \citep{PeerIndex} that market and sell to
businesses and other organizations influence scoring metrics.  The
most standard metric employed is the number of followers an account
has. However, a number of studies \citep{cha, weng} have concluded that
it is not a good indicator, since most followers fail to engage with
the messages that have been broadcast.  For that reason, the number of
retweets an account receives \citep{kwak} is a better measure of
influence. Since we are interested in ranking of users, more
sophisticated influence measures based on the popular PageRank
\citep{page1999pagerank} and HITS \citep{kleinberg-authority} ranking
algorithms, widely used for ranking search results on the Web, have
been used \citep{kwak, gayo-avello}. However, these algorithms have
been applied to the followers network. The latter clearly captures the
popularity of users, but not necessarily of their influence. For
example, the top twenty most followed accounts with a minimum of 25
million followers comprise of entertainers and athletes, the sole
exception being President Obama.

In this paper, we propose to measure an account/user's influence on
the Twitter social media platform, by taking into consideration both
their ability to produce new content by posting messages, but also to
generate interactions from other accounts through retweeting and
mentioning. To that end, we build a statistical model for an account's
actions and interactions with other accounts.  It uses a counting
process framework to capture the posting, retweeting and mentioning
actions. In addition, based on this model we introduce a novel {\em
  influence measure} that leverages both the follower network (that
captures the potential for posted messages to generate interactions
with other users) and the {\em intensity} over time of the basic
actions involved (posting, retweeting and mentioning).  Hence,
underlying the model in this paper is the idea that conversations, and
in particular the rate of directed activity, between accounts reveal
their real-world position and influence.

We illustrate the model on a closely knit community, namely that of
the members of the United States Senate, the upper legislative house
in the bicameral legislative body for the United States. Two senators
are democratically elected to represent each state for six year terms.
We further augment the set of Twitter accounts analyzed by including
selected prominent news organizations (e.g. Financial Times,
Washington Post, CNN), as well as popular bloggers (e.g. Nate Silver,
Ezra Klein), the accounts of President Obama and the White
House, and two influential federal agencies (the US Army and the
Federal Reserve Board); for details refer to Section
\ref{senate-results}.

\begin{figure}[!t]
\includegraphics[width=1\columnwidth]{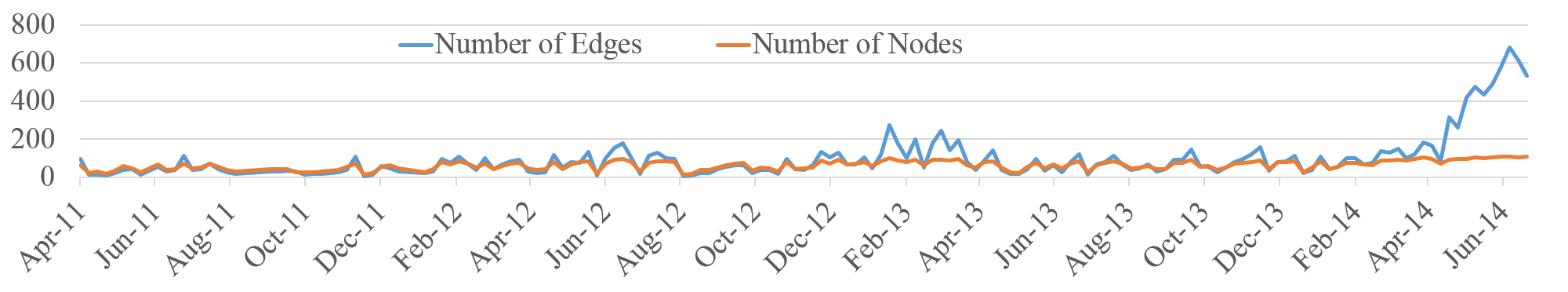} 
\includegraphics[width=0.24\columnwidth,trim=2.5cm 2cm 1.8cm 1.2cm, clip=true]{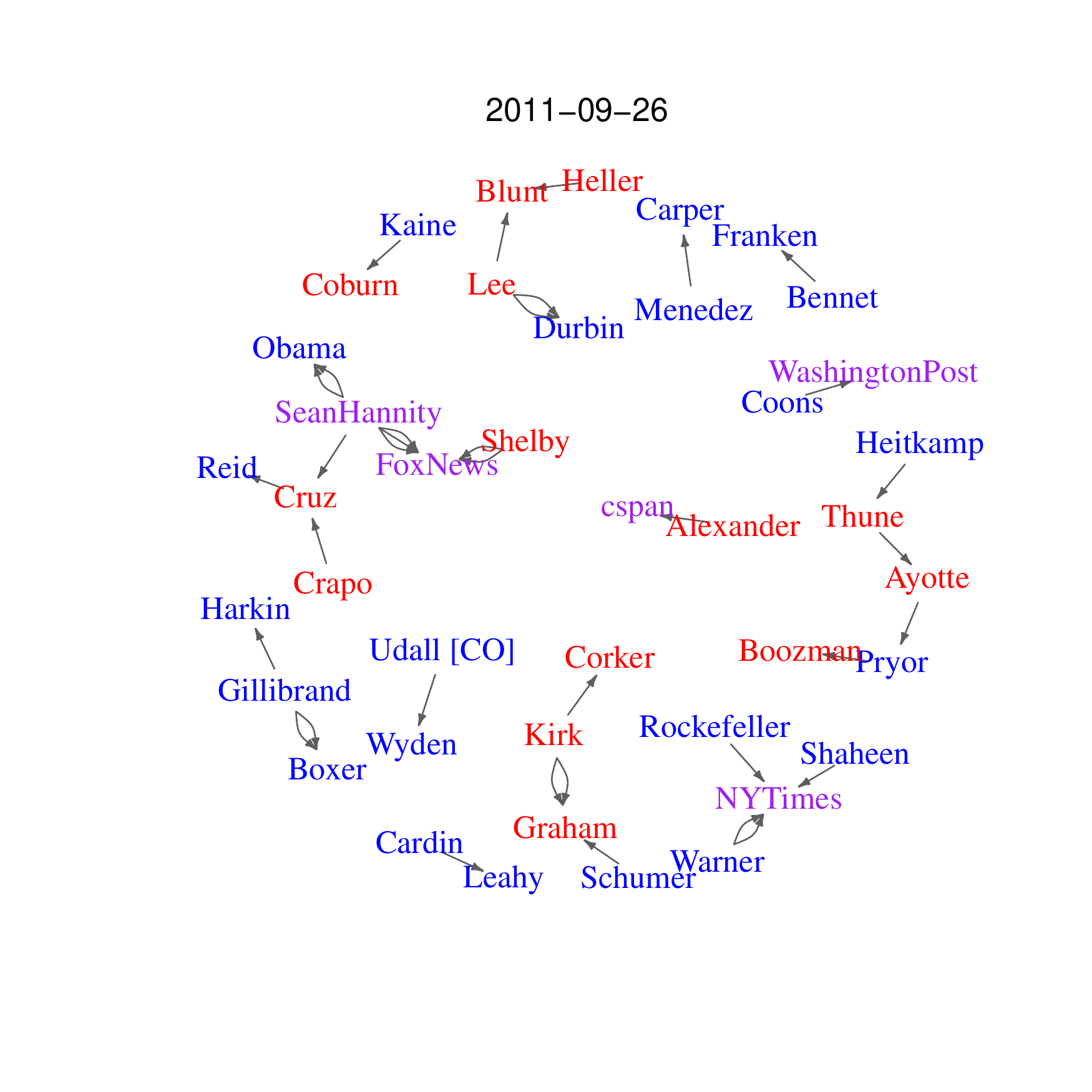}
\includegraphics[width=0.25\columnwidth,trim=2.5cm 2cm 1.8cm 1.2cm, clip=true]{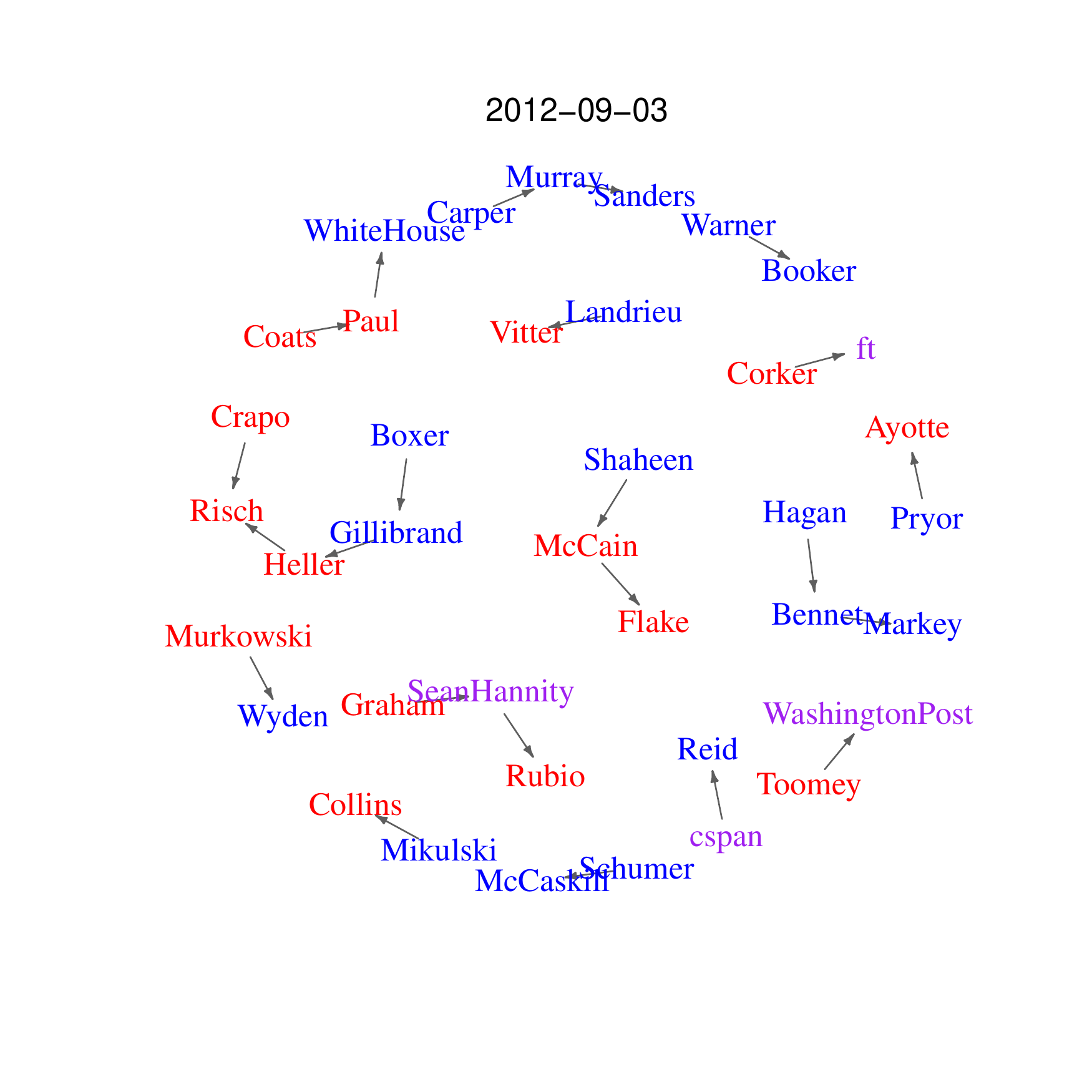}
\includegraphics[width=0.24\columnwidth,trim=2.5cm 1.8cm 1.8cm 1.2cm, clip=true]{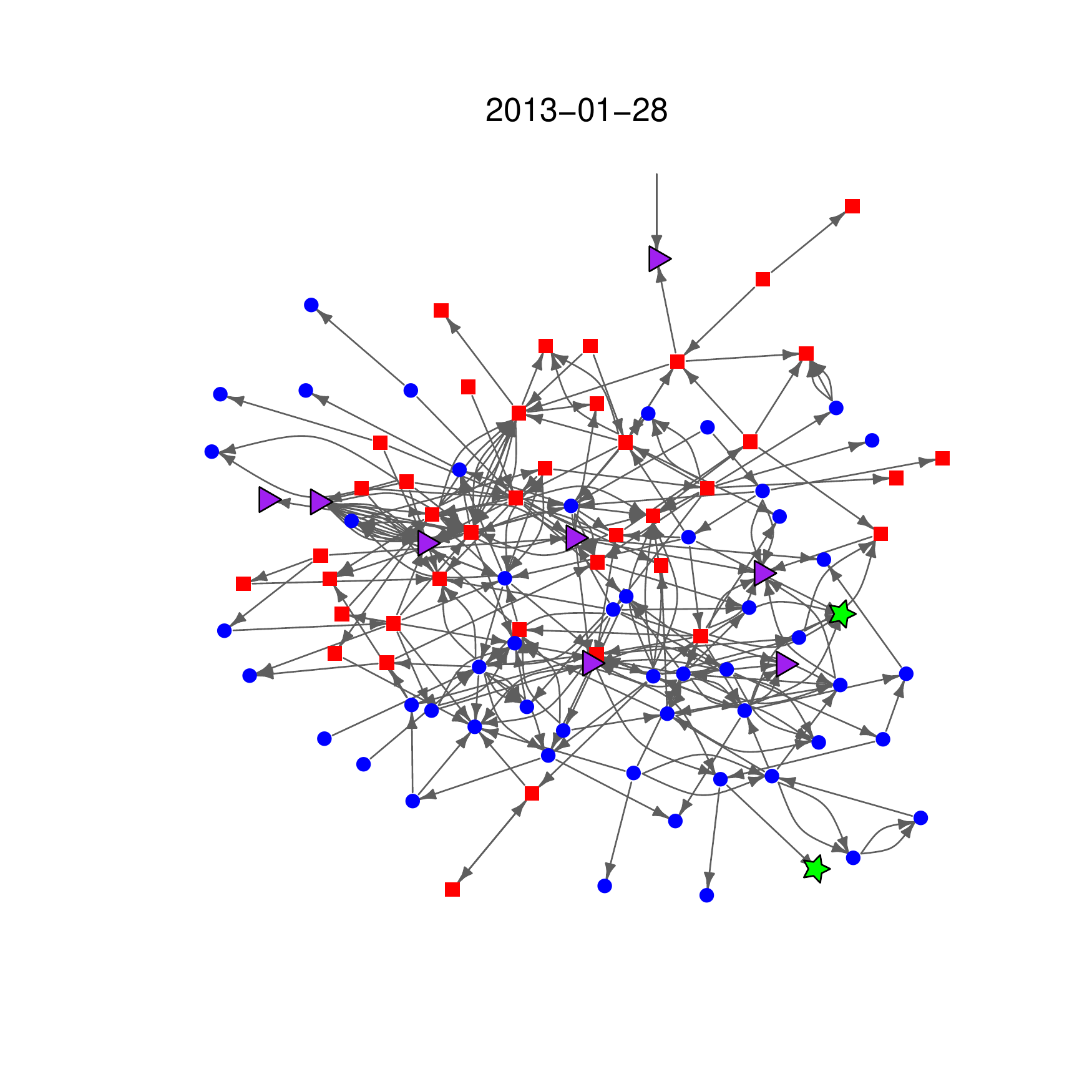}
\includegraphics[width=0.24\columnwidth,trim=5.5cm 5.1cm 1.8cm 1.2cm, clip=true]{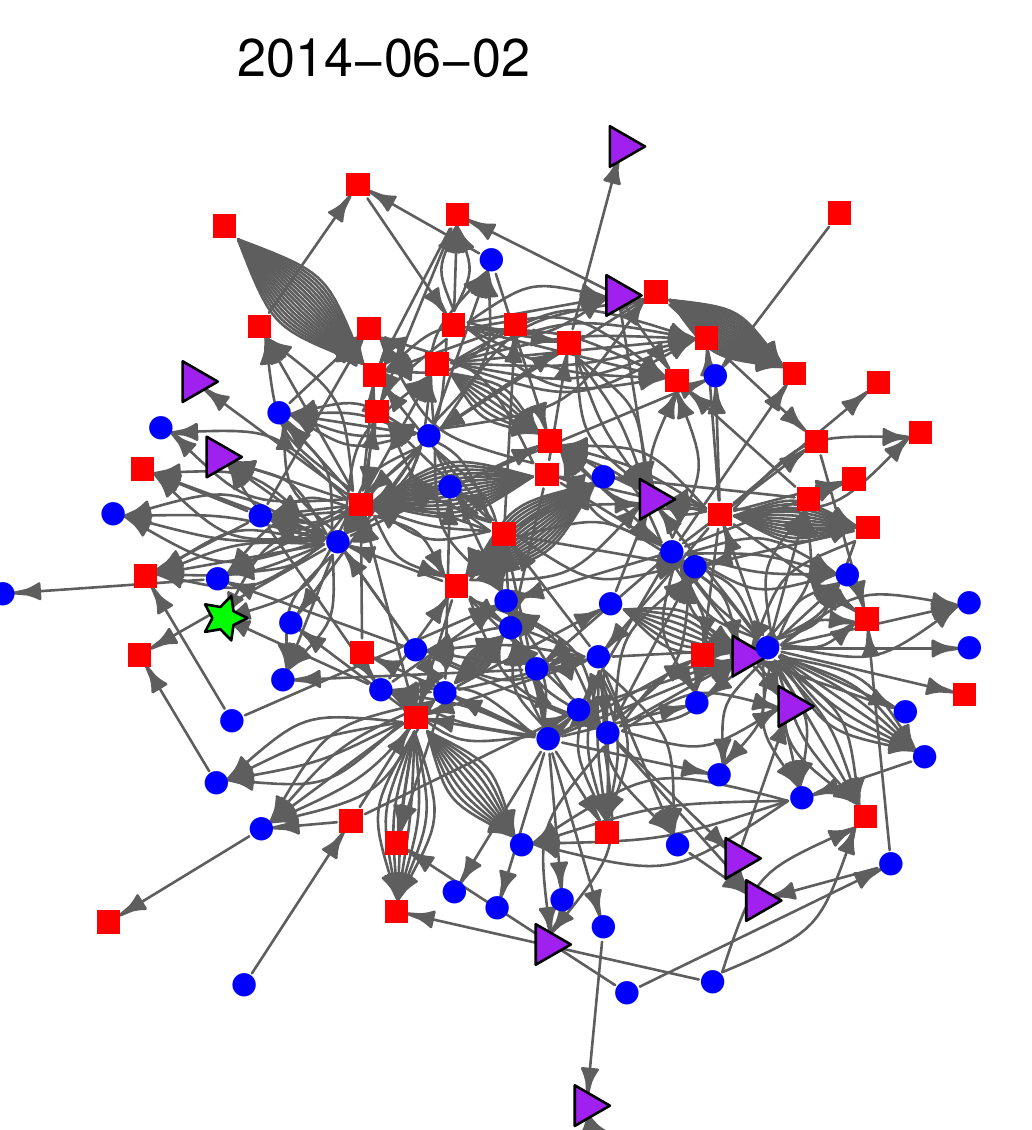}
\caption{Weekly Twitter (mentions and retweet) network statistic time-series and drawings. The nodes (Twitter accounts) contain democratic senators (blue circles), republican senators (red squares), media (purple triangles), and government agencies (green stars).}
\label{fig:network_drawings}
\end{figure}

The retweeting and mentions interactions from our data are drawn as
directed edges in Figure~\ref{fig:network_drawings}. Given this sequence 
of network snapshots, we identify particular senators and news agencies that tend 
to elicit interactions from other accounts (i.e., have many incoming edges relative 
to how often they tweet), thus revealing their influence on Twitter. 
Our results further indicate \ref{senate-results} that the proposed approach produces
influence measures for the U.S. Senators that correspond more closely 
with their legislative importance than purely network-based solutions
based on the PageRank algorithm.


The remainder of the paper is organized as follows: in Section 2, we
introduce the modeling framework and the proposed influence
measure. Section 3 presents the algorithm to obtain the model
parameter estimates, as well as establish their statistical properties
and those of the influence measure. The performance of the model is
evaluated on synthetic data sets in Section 4, while the US Senate
application is presented in Section 5. Finally, some concluding
remarks are drawn in Section 6.

\section{The model and the influence measure}\label{model}

We start our presentation by defining some key quantities for future developments. Let $G=(V, L)$ denote the followers network, where 
$V$ corresponds to the set of nodes of all the Twitter accounts under consideration and $L=\{L_{i,j}, 1\leq i \neq j\leq n\}$ the edge set
between them and captures whether an account foillows another account..
Note that the network is bidirectional in nature and not necessarily symmetric, since account $i$ may follow account $j$, but not vice versa.
In principle, $L$ can be dynamically evolving, but in this work we consider $L$ to be static and not changing over time.
As explained in the introductory Section, in the Twitter platform, accounts (nodes) can undertake the following three actions: post a new message,
retweet a message posted by another account that they follow and finally mention another account that they follow. Further,
the vast majority of messages posted, retweeted or mentioned have key terms (wth a $\sharp$ prefix) that identify the topic(s) 
that are been discussed. 

Next, we define the following two key counting processes. Let $N_j(t)$ denote the total number of retweets and mentions that account $j$ generates on topic $l$ by time $t$ and by 
$A_j(t,l)$ the total number of posted messages by account $j$ on topic $l$ by time $t$. Define $\alpha_j$ to be a parameter that captures the long-term capability of account $j$ to
generate responses by other accounts from the content posted, and $\beta_j$ a parameter that captures the long term susceptibility of account $j$ to respond (retweet/mention) to the postings of
the accounts it follows. We model $\{N_j(t,l)\}_{i=1}^n$ as a set of counting processes through their hazard rates, using a version of Cox \citep{andersen1982cox} proportional hazard model; specifically,
the hazard rate $\lambda_{j,l}(t)$ of process $N_j(t,l)$ is given by 
\begin{equation}
\lambda_{j,l}(t)=\lambda_{0,l}(t)\exp\left(\sum_{i\neq j}L_{ij}(\alpha_i+\beta_j) \log(M_i(t,l)+1)\right), \label{RMH}
\end{equation}
where $M_j(t,l)=A_j(t,l)+N_j(t,l)$ the total number of posting, retweets and mentions for account $j$ on topic $l$ by time $t$.
We assume that the parameters  $\alpha_i,\beta_i\in (-\infty, \infty)$,  since accounts and their users may be positively or negatively inclined towards other accounts, as well as being more keen in
joining specific conversations or passively retweeting messages from favorite accounts. The nonparametric baseline component $\lambda_{0,l}(t)$ is time varying. In general, we would expect this
baseline to be small for large times $t$, since topics in social media platforms have a high churn rate; they become "hot" and generate a lot of action over short time scales and afte awhile it stops being
discussed. The model posits that account $j$ interacts with other accounts at a baseline level $\lambda_{0,l}(t)$, modulated by its ability to generate responses by accounts in its followers network, as well as
its own susceptibility to respond to accounts it follows postings and rebroadcasting of messages. Note that we model the retweet-mention process $N_j(t,l)$, since it reflects interactions between nodes and
use the total activity process $M_j(t,l)$ as a covariate.

To complete the modeling framework, denote the set of topics in the data as $\{1,\ldots, \Gamma\}$. Further, let $\mathcal{T}_j^l=\left\{T_{j,1}^l,\ldots, T_{j,n_j}^l\right\}, t=1,\cdots,n_j,$ denote the 
set of time points that account $j$ took action (post, retweet, mention) on topic $l$. Finally, for identificaiton purposes, we require one member of the parameter vector 
$\Omega=(\alpha_1,\alpha_2,\ldots, \alpha_n, \beta_1,\ldots,\beta_n)$ to be set to a fixed value, and without loss of generailty we set $\alpha_1=0$. Following, \citep{andersen1982cox}, we employ a
partial-likelihood function to obtain estimates of $\Omega$. Specificially, we treat the baseline $\lambda_{0,l}(t)$ as a nuisance parameter and decomposing the full-likelihood to obtain
\begin{equation*}
PL=\prod_{1\leq l\leq \Gamma} \left( \prod_{1\leq j\leq n}\prod_{1\leq k \leq n_j}\frac{\lambda_j(T_{j,k}^l)}{\sum_{1\leq i\leq n} \lambda_i(T_{j,k}^l)}
\right)
\end{equation*}
 Plugging the exact form of the hazard rate from (\ref{RMH}) into the partial-likelihood function (PL), we get:
\begin{equation}
  \begin{aligned}
PL=\prod_{1\leq l\leq \Gamma} &\left( \prod_{1\leq j\leq n}\prod_{1\leq k \leq n_j}\frac{\exp\left(\sum_{i\neq j}L_{ij}(\alpha_i+\beta_j) \log(M_i(T_{j,k}^l,l)+1)\right)}{\sum_{1\leq i\leq n} \exp\left(\sum_{u\neq i}L_{ui}(\alpha_u+\beta_i) \log(M_u(T_{j,k}^l,l)+1)\right)}\right)
\label{PL}
  \end{aligned}
\end{equation}

\subsection{The Influence Measure}

Leveraging the structure of the model, we propose to measure an account's (node) influence as  the total hazard rate change it will bring to its followers.
Specifically, for an account $j$  its hazard rate at time $t$ is given by:
$H_j=\exp\left(\sum_{k\neq j}\log (M_k(t)+1) L_{kj}(\alpha_k+\beta_j)\right)$. Further, the contribution of node $i$ is 
$H_j^{(i)}=\exp\left(\log (M_i(t)+1) L_{ij}(\alpha_i+\beta_j)\right)$.
Then, after some algebra we obtain that the total hazard rate change $i$ brings to its followers can be written as:
\begin{equation}\label{influence.10}
TH^{(i)}=\sum_{j\neq i} L_{ij}\cdot \exp\left(\log (M_i(t)+1) (\alpha_i+\beta_j)\right).
\end{equation}
Since $M_i(t)$ is a random value, we approximate it by its observed average value, $\bar{M}_i$, calculated from the data. 
Hence, the influence measure becomes
\begin{equation}\label{influence.20}
\tilde{TH}^{(i)}=\sum_{j\neq i} L_{ij}\cdot \exp\left(\log (\bar{M}_i+1) (\alpha_i+\beta_j)\right).
\end{equation}
Finally, we express it in a log-scale, so as to linearize the scale and make it compatible with the range of values of the response and susceptibility parameters $\alpha$ and $\beta$:
\begin{equation}\label{influence.30}
\Xi^{(i)}=\log\left[\sum_{j\neq i} L_{ij}\cdot \exp\left(\log (\bar{M}_i+1) (\alpha_i+\beta_j)\right)\right].
\end{equation}

\section{Computation and Inference}

Next, we present a Newton-type algorithm for computing the parameter estimates $\Omega$. The objective function corresponds to the logarithm of the partial likelihood
function  (\ref{PL}) given by
\begin{equation}
  \begin{aligned}
LL=\log(PL)=\sum_{1\leq l\leq \Gamma} &\left\{ \sum_{1\leq j\leq n}\sum_{1\leq k \leq n_j}\sum_{i\neq j}L_{ij}(\alpha_i+\beta_j) \log(M_i(T_{j,k}^l,l)+1)\right.\\
&-\sum_{1\leq j\leq n}\sum_{1\leq k \leq n_j}\log \left[\sum_{1\leq i\leq n} \exp\left(\sum_{u\neq i}L_{ui}(\alpha_u+\beta_i) \log(M_u(T_{j,k}^l,l)+1) \right)\right]\\ \label{LL}
  \end{aligned}
\end{equation}
Due to its smoothness we employ Newton's algorithm that uses the gradient and the Hessian of $LL$. The detailed expressions
for the gradient vector $G\equiv\nabla_\Omega LL$ and the Hessian $H\equiv \nabla_\Omega \nabla_\Omega (LL)$ are given in the Appendix.

%

\begin{algorithm}
 \begin{algorithmic}[1]
 \STATE Initialize the vector $\Omega$ value by $\alpha_1=\ldots=\alpha_n=\beta_1\ldots=\beta_n=c= 0$ 
 \STATE{Define $s$ as a positive thresholding constant for the minimum step size}
 \WHILE{$t>s$} 
 \STATE Calculate $G$ by using (\ref{PA}) and (\ref{PB})
 \STATE Calculate $H$ by using (\ref{PAE}) to (\ref{PABD})
 \STATE Find the optimum positive $\tau$ value such that $\Omega-\tau\cdot H^{-1} G$ will maximize the log-partial-likelihood (\ref{LL})
 \STATE  Update $\Omega\leftarrow\Omega-\tau\cdot H^{-1} G$.
 \STATE In the updated $\Omega$, set $\alpha_1=0$.
 \ENDWHILE
 \RETURN $\Omega$
 \end{algorithmic}
 \caption{Estimating the parameters by Newton's algorithm}
 \label{newton}
\end{algorithm}

The steps of the optimization are given in Algorithm~1.
As stated in the algorithm,  $s$  is a positive constant to judge the convergence of the the Newton's algorithm. The computational complexity of this algorithm is 
dominated by the computation of $H$. Denote by $m_n=\max_{1\leq j\leq n}\{n_j\}$.   Based on (\ref{PA}) and (\ref{PB}),  it costs $O(\Gamma n m_n)$ 
operations to calculate an entry of $G$. Further, siince $G$ is of dimension $2n$, it takes $O(\Gamma n^2 m_n)$ to obtain the entire $G$ vector. 
Analogously, based on (\ref{PAE}) to (\ref{PABD}),  it costs $O(\Gamma n m_n)$ 
operations to calculate an entry of $H$, if proper book-keeping is used on the results obtained for the gradient $G$. 
Further, since $H$ is of dimension $n^2$, it takes $O(\Gamma n^3 m_n)$ to obtain the entire $H$ matrix. 
Hence, the overall time complexity for each iteration of the algorithm is of the order $O(\max\{\Gamma n^3 m_n\})$.

\section{Properties of the $\hat \Omega$ estimates}

Next, we establish that the estimator $\hat{\Omega}$ which maximizes (\ref{LL}) will converge to the true parameter 
$\Omega$ in probability in probability under certain regularity conditions. Before we state the main result, we present some definitions. Let
\begin{equation*}
  \begin{aligned}
& E_{j}(t,\Omega)=\Gamma^{-1}\sum_l \lambda_{j,l}(t)=\Gamma^{-1}\sum_l  \lambda_{0,l} \exp\left(\sum_{i, i\neq j}L_{ij}(\alpha_i+\beta_j) \log(M_i(t,l)+1) \right)\\
& \Phi_j=(\phi_1,\ldots,\phi_n):=(\alpha_1+\beta_j,\ldots, \alpha_n+\beta_j)\\
& \Phi_j'=(\phi_1',\ldots, \phi_n'):=(\alpha_1'+\beta_j' ,\ldots, \alpha_n'+\beta_j' )' \\
& E_{j}^{(1)}(t,\Omega)=\left(\frac{\partial E_{j}(t,\Omega)}{\partial \phi_1},\dots, \frac{\partial E_{j}(t,\Omega)}{\partial \phi_n}\right)\\
& E_{j}^{(1)}(t,\Omega')=\left(\frac{\partial E_{j}(t,\Omega')}{\partial \phi_1'},\dots, \frac{\partial E_{j}(t,\Omega')}{\partial \phi_n'}\right)\\
  \end{aligned}
\end{equation*}

Then, we can establish the following result.

\textbf{Theorem 1.}
Suppose there exists time point $t_0$, such that all the observed time points 
satisfy $T_{j,k}^k\leq t_0$ . Further, for each topic $l$,  the observation times  
$T_{j,k}^l,1\leq k\leq n_j$ are different, for all $1\leq l\leq \Gamma$. (Note however, that for different topics we allow overlap of event occuring times.) 
Further, we assume that 

(A) (Finite interval) $\int_0^{t_0}\lambda_0(u)du<\infty$.

(B) There exists $e(\Omega',t)$ such that
\begin{equation*}
  \begin{aligned}
& \sup_{ t\in[0,t_0],\Omega'}\sum_j |{E}_{j}(t,\Omega')-{e}(\Omega',t)|\to_P 0' \\
  \end{aligned}
\end{equation*}
And if we denote 
\begin{equation*}
  \begin{aligned}
& e_{j}^{(1)}(t,\Omega')=\left(\frac{\partial e_{j}(t,\Omega')}{\partial \phi_1'},\dots, \frac{\partial e_{j}(t,\Omega')}{\partial \phi_n'}\right)\\
& e_{j}^{(2)}(t,\Omega')=\left(\frac{\partial^2 e_{j}(t,\Omega')}{\partial \phi_i' \partial \phi_j' }\right)\\
  \end{aligned}
\end{equation*}
then also,
\[\sup_{ t\in[0,t_0],\Omega'}\sum_j \|{E}_{j}^{(1)}(t,\Omega')-{e}^{(1)}(\Omega',t)\|\to_P 0\]
\[\sup_{ t\in[0,t_0],\Omega'}\sum_j \|{E}_{j}^{(2)}(t,\Omega')-{e}^{(2)}(\Omega',t)\|\to_P 0\]
where again $\|\cdot\|$ is the $L_2$ norm of a vector or matrix. 

(C)  $e(\Omega',t)$ is bounded away from zero. $e(\Omega',t)$ and $e^{(1)}(\Omega',t)$ are continuous functions of $\Omega'$, uniformly in $t\in [0,t_0]$. $e^{(2)}(\Omega,t)$ is positive definite. 

Then, under condistions (A-C), we have that  
$$\hat{\Omega}\to_P \Omega.$$ 

The detailed proof is given  in the Appendix.

Based on Theorem 1, by leveraging the properties of continuous functions, we can establish
the consistency of the proposed inflience measure.

\textbf{Proposition 1}. Let  $\Xi(t)=(\Xi^1(t), \cdots, \Xi^n(t))$ denote  the $n$-dimensional vector of influence measures at time $t$. Further, denote by
$\hat{\Xi}(t)=(\hat{\Xi}^1(t), \cdots, \hat{\Xi}^n(t))$  their empirical estimates.  Under the conditions of Theorem 1, we have that
\begin{equation}
\left\|\hat{\Xi}(t)-\Xi(t)\right\|\to_P 0
\end{equation}
for any $t\geq0$. 

Based on Theorem 1, the proof of the proposition is straightforward, since each element of the vector $\hat{\Xi}$ is a continuous function of $\hat{\Omega}$.

\section{Performance evaluation}

In this section, we evaluate the proposed model and influence measure on synthetic data. We start by outlining the data generation mechanism.

\noindent
\textbf{Step 1}: Building the followers network $L$. \\
The steps employed for this task are presented next.
\begin{itemize}
\item
 First, for each node $i$, generate $K_1(i)$ from a uniform distribution on the integers $\{1,\ldots, K\}$, where $K=\lfloor*n/2 \rfloor$ and $\lfloor*\cdot\rfloor$ is the floor function that returns the maximum integer not larger than the value inside.
\item
Generate $F_1(i)$ for node $i$  by randomly sampling $K_1(i)$ users from $\{1,\ldots,n\}\backslash\{ i\}$ . If $k \in F_1(i)$, let $L_{ik}=1$, $ 1\leq i\leq n$.
\item
For each node $j$, sample $K_2(j)$ uniformly from the set $\{1,\ldots, K\}$.  Generate $F_2(i)$ for node $j$  by randomly sampling $K_2(j)$ users
from $\{1,\ldots,n\}\backslash\{ j\}$ . If $k \in F_2(j)$, let $L_{kj}=1$, $ 1\leq j\leq n$.
\end{itemize}
At the end of this procedure,  every node in the network has at least one follower and at least an account that it follows.\\
\noindent
\textbf{Step 2}. Generate the post, retweets and mentions sequences.\\
Since the baseline hazard rate $\lambda_{0,l}(t)$ always gets canceled in the partial- likelihood function (\ref{PL}), we select  $\lambda_{0,l}(t)$ as $\lambda_{0,l}(t)=a$, whenever $0\leq t\leq t_0$ and $\lambda_{0,l}(t)=0$ when $t>t_0$, where  $a$ is a small positive constant.

We then generate actions on this network withthe following algorithm  below for each topic $l \in \{1,\ldots,\Gamma\}$. In this algorithm, we let the nodes send tweets with probability $p$ at each time point $0,1,\ldots, \lfloor*t_0\rfloor$, while we generate the retweets and mentions in the standard survial analysis way, by using the hazard rate (\ref{RMH}).

{\tiny

\begin{algorithm}
 \begin{algorithmic}[1]
 \STATE Initialize Indicator which is the sequence to record the nodes that have  mentioned or retweeted as an empty sequence. 
\STATE Initial t=0. Let each node has a tweet with probability $p$.
\STATE Let each node send out tweets from Binomial$(J,p)$.
 \WHILE{$t<t_0$ (stopping time for all topics)}
 \STATE Generate survival time for each node with its hazard rate (\ref{RMH})
 \STATE Find node $i$ with the shortest time $t_s$. 
\IF {$t+t_s<t_0$}
	\STATE Update $t$ to be $t+t_s$. Record the node that has done this retweet or mention. 
\ENDIF
\IF {$t+t_s>t_0$}
	\STATE Break
\ENDIF
\ENDWHILE
 \RETURN Indicator
 \end{algorithmic}
 \caption{Generate Group A actions}
 \label{algo:actions}
\end{algorithm}
}

We first illustrate the performance of the Newton estimation algorithm, on a random network of varying size. We set the parameter $a=0.5$ for the
baseline hazard rate and choose a time horizon of $t_0=7$, to emulate a week's worth of data.  We also select the parameters $\Omega$ uniformly at random in the interval  $[-0.3,0.3]$. With different network sizes $n$ and number of topics generated $\Gamma$, we obtain the following two plots to show the mean squared error of the parameter and influence estimates $\frac{\|\hat{\Omega}-\Omega\|}{\sqrt{2n-1}}$ and $\frac{\|\hat{\Xi}-\Xi\|}{\sqrt{n}}$, where $\|\cdot\|$ corresponds to the $\ell_2$ norm of a vector. The results are based on 20 replicates of the underlying followers networks, as well as
the actions (postings, retweets and mentions) data.

\begin{figure}[ht]
\centering
 \includegraphics[width=.49\linewidth]{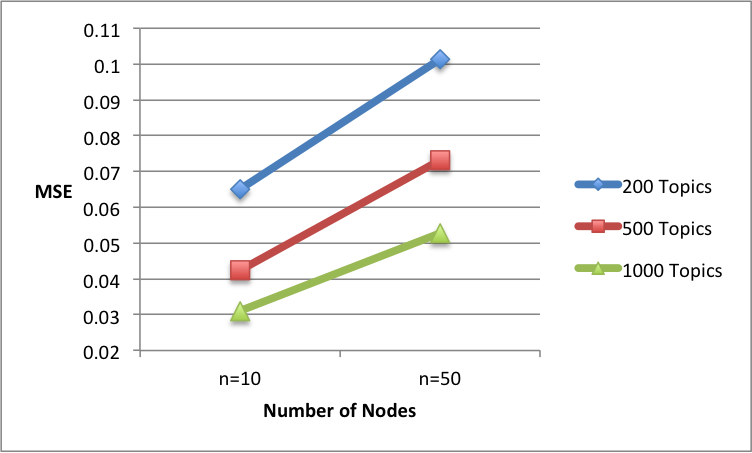}
 \includegraphics[width=.49\linewidth]{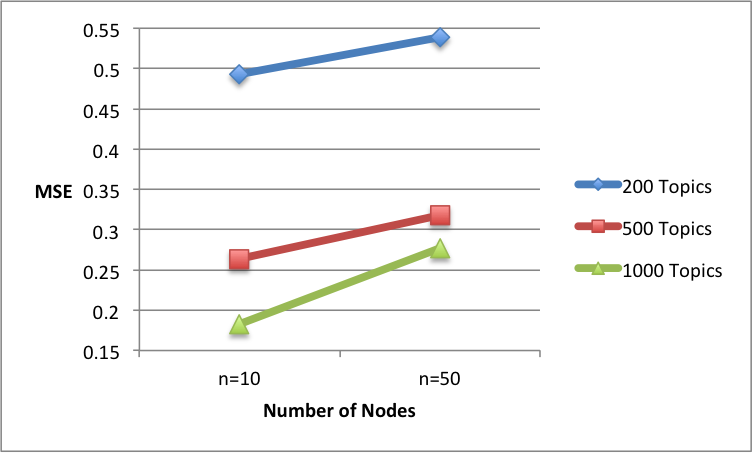}
\caption{Mean squared error of the model parameter estimates $\Omega$ (left) and  $\Xi$ (right).}
\end{figure}


It can be seen that the quality of the estimates improves as a function of the number $\Gamma$ of topics discussed, while it deteriorates 
as a function of the number of nodes in the followers network $L$. Another way to look at the quality of the estimates, is
to examine the relative error of the parameter and influence estimates, given by $\frac{\|\hat{\Omega}-\Omega\|}{\|\Omega\|}$ and $\frac{\|\hat{\Xi}-\Xi\|}{\|\Xi\|}$.

It can be seen in the following Figure that especially the
influence measure which is of prime interest in applications, exhibits a small (less than 10\%) relative error rate.

\begin{figure}[ht]
\centering
  \centering
 \includegraphics[width=.49\linewidth]{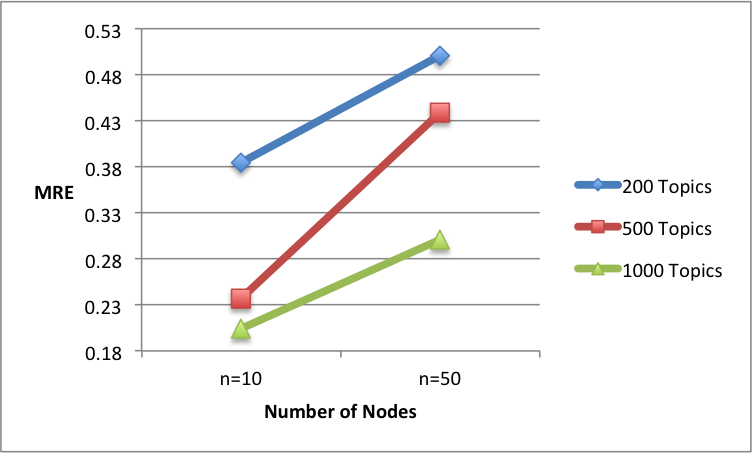}
 \includegraphics[width=.49\linewidth]{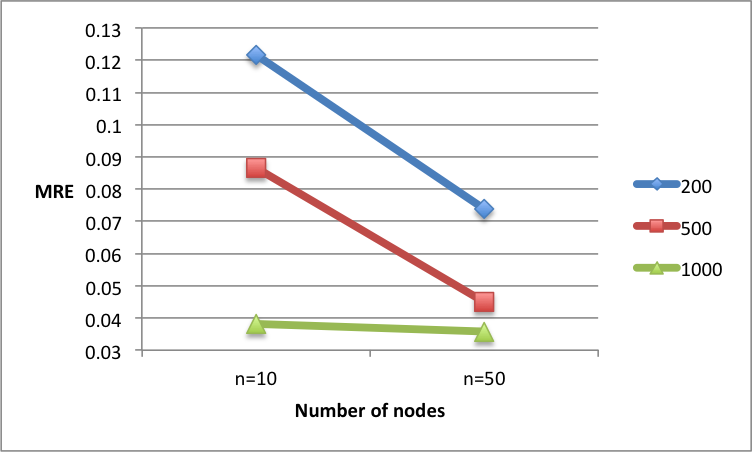}
\caption{Mean relative error of the model parameter estimates $\Omega$ (left) and $\Xi$ (right).}
\end{figure}

\begin{figure}[htbp]
\centering
\includegraphics[scale=0.6]{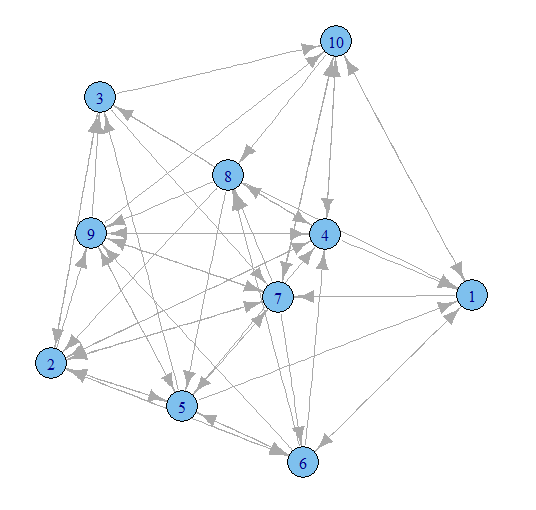}\label{artifical-topology}
\end{figure}

Next, we use a size 10 network, specially constructed to gain insight into the workings of the proposed influence measure.
The settings for the data generation are as follows: $\Gamma=500$, $\alpha_1=0, \alpha_2=-2, \alpha_3=\cdots=\alpha_{10}=0.2$ and $\beta_{1}=\cdots=\beta_{10}=0$. Finally, the topology of the followers networks is given in Figure \ref{artifical-topology}.

\begin{figure}[htbp]
\centering
\includegraphics[scale=0.6]{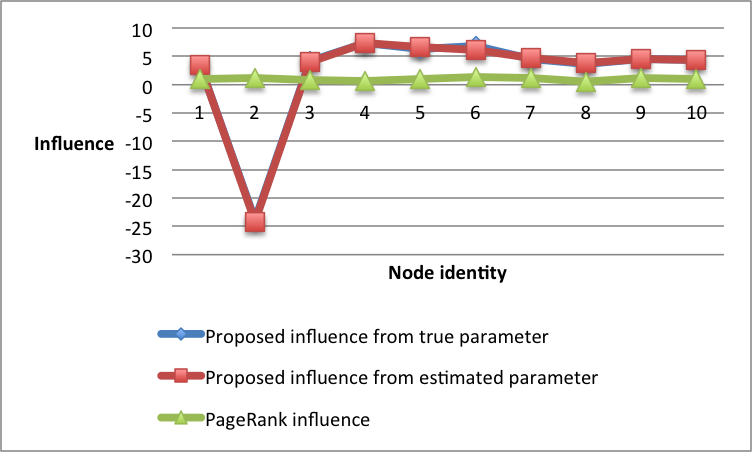}\label{results-art-topology}
\end{figure}

Since $\alpha_2=-2$, node 2 is an ``unpopular" one and hence can hardly generate any retweets and mentions of its postings. On the other hand, 
all nodes have approximately an equal number of followers, which suggests that their ranking according to the PageRank metric (or many other popular
ones based on that network) will be approximately similar. The results based on a single realization of the user actions data generation process is shown
iin Figure \ref{results-art-topology}. It can be seen that relying on the followers network structure gives a false impression, while the proposed 
influence measure that incorporates the actions of the accounts provides a more insightful picture.

\section{Identifying Influential Senators} \label{senate-results}
Tweets and follower lists are collected using Twitter's API and consist of approximately
200,000 tweets and 4671 follower links within the set of 120 accounts from April 2009 to July 2014. 
The retweeting and mentions interactions are drawn in Figure~\ref{fig:network_drawings}, where 
accounts are registered to 55 Democratic politicians (U.S. Senators and the President of the U.S.), 
46 Republican Senators, 2 government organizations (U.S. Army and the Federal Reserve Board), and 
16 media outlets, including newspapers (Financial Times, Washington Post, New York Times, Huffington Post), 
television networks (MSNBC, Fox News, CNN, CSPAN), reporters (Nate Silver (538), Ezra Klein) and 
television hosts (Bill O'Reilly, Sean Hannity). 
The figure shows some periods of increased activity, as in the months surrounding the inauguration of 
President Obama (January 2013), the debate on raising the debt ceiling of the US government and its temporary suspension
around April 2013 and the summer of 2014 (soccer World Cup). Note that the sudden increase during the summer of 
2014 may be an artifact of rate limiting data acquisition. Specifically, Twitter's API allows access to only the past $3000$ tweets for any account. 
As a consequence, for extremely high volume users, like newspapers and television networks, 
our data traces their Twitter usage for months. For the least active users in our data, 3000 tweets 
dates back multiple years. 

An inspection of actual tweets in Table~\ref{table:tweets} shows, consistent with \cite{twitterCongress}, 
that senators tend to retweet and mention as a means of self or legislative promotion. 
In fact, we see a number of references to legislative activity, such as calls for gun reform, carbon emissions, and
 references to actual bills on overtime pay, domestic violence protections, among others. Senators often cite news 
coverage by retweeting or mentioning news media accounts that support their political agenda, which would suggest
 that the media outlets collectively have enormous influence. This also suggests that Twitter is utilized by senators as
part of a larger strategy to build and coalesce public support in order to pass bills through congress. 

\begin{table}
\centering
\caption{Actual tweets mentioning or retweeting the most influential accounts over from May 15, 2014 to July 3, 2014.}
\label{table:tweets}
\begin{tabular}{|l|c|p{4in}|}\hline
Date & Account & Tweet  \\\hline\hline
05/19/2014 & Menendez & ``.$@$SenBlumenthal \& in \#NJ the avg student loan debt is over \$29K. It's unacceptable! \#GameofLoans http://t.co/hUJMSeJbfd''\\
05/23/2014 & Cornyn & ``RT @nytimes: Former Defense Secretary Gates Is Elected President of the Boy Scouts http://t.co/C7STUSVIP3''\\
05/27/2104 & Blumenthal  & ``RT $@$msnbc: $@$SenBlumenthal calls for reviving gun reform debate after mass shooting near Santa Barbara: http://t.co/7sqtf1IAFy''\\
06/02/2014 & Markey & ``RT $@$washingtonpost: A huge majority of Americans support regulating carbon from power plants  http://t.co/lj6ieL5D1Y http://t.co/2CA63hTqmm''\\
06/17/2014 & Markey & ``Proud to intro new bill w $@$SenBlumenthal 2 protect domestic violence victims from \#gunviolence http://t.co/MsgK40oLiT http://t.co/ynEHrEbh2x''\\
06/20/2014 & Blumenthal & ``Proud to stand w/ $@$CoryBooker \& others on enhancing rules to reduce truck driver fatigue. Their safety \& safety of others is paramount. -RB''\\
06/20/2014 & Markey & ``Proud to support our workers and this commonsense bill w $@$SenatorHarkin Keeping Track: Overtime Pay, via $@$nytimes http://t.co/TnAS96Hro5''\\
06/25/2014 & Durbin & ``Watch now: $@$OfficialCBC $@$HispanicCaucus $@$CAPAC $@$USProgressives $@$SenatorCardin on racial profiling \#MoreThanAProfile http://t.co/ZX0Eu65dgi\\
06/25/2014 & Cardin & ``RT $@$TheTRCP: Thank you $@$SenatorCardin for standing with sportsmen today for \#CleanWater \#protectcleanwater''\\
06/27/2014 & Markey & ``Thanks $@$alfranken $@$CoryBooker $@$amyklobuchar $@$SenBlumenthal for joining me in support of community \#broadband http://t.co/O8Px2MzrCg''\\
06/27/2014 & Menendez & ``Took my first \#selfie at \#NJs $@$ALJBS! Hope $@$CoryBooker is proud of his NJ Sen colleague. http://t.co/FrEJonUy9d\\
06/28/2014 & Booker & ``Thanks Adam RT $@$AIsaacs7 Props to $@$CoryBooker and $@$SenRandPaul for their bipartisanship in introducing their amendment \#MedicalMarijuana''\\\hline
\end{tabular}
\end{table}

To test these hypotheses and also rigorously compare the proposed influence measure to PageRank applied to the followers networks (which constitutes the backbone
of many ranking algorithms of Twitter accounts), 
we perform a regression analysis to assess how well each measure explains  
{\em legislative leadership} in Congress. Our response variable is the leadership score, 
published by \url{www.govtrack.us} \citep{GovTrack}. 
GovTrack creates the leadership score by applying the PageRank algorithm to the adjacency matrix 
of bill cosponsorship data. Thus, the leadership score for each senator is a number between 
$0$ and $1$, where higher values denote greater legislative leadership. The regression model 
we are interested in is 
\begin{equation} \label{eqn:realdataRegn}
\text{Leadership} = \beta \text{Influence} + \Theta \text{Controls}, 
\end{equation}
where Influence contains the proposed measure and/or PageRank, and Controls includes party affiliation, 
gender, age, and number of years in the senate. Seniority endows a number of benefits including preferential 
assignment to committees. Thus, these control variables likely associate strongly with legislative leadership. 

To estimate the proposed influence measure, the data is organized into weekly intervals after using the follow-follower relations to construct the adjacency matrix $L$.
In Twitter it is common to use ``hashtags'' or the \# symbol followed by a user-specified category to identify context, which, as mentioned in Section 1 can be used as an 
indicator of different conversations. However, we find that senators do not utilize hashtags often. To overcome this challenge, we follow previous works on Twitter
 \citep{hong2010empirical, lda_tweets} by applying probabilistic topic modeling, which was first introduced in \cite{blei2003latent}. Extensive work in computer science and
 applied statistics has led to fast algorithms capable of analyzing extremely big text archives.  Due to space constraints, for 
statistical and algorithmic details on the topic model,  see \cite{blei2012probabilistic,blei2007correlated} and references therein.

Topic modeling applied to the data results in a soft clustering of tweets into groups (topics), which is appropriate since a single tweet could touch on multiple issues. 
Thus, tweets are assigned to topics that had at least $0.25$ probability. Given the fast moving landscape of social media, new topics are assigned each week,
 leading to $2770$ topics in total for the entire data set. After preprocessing, we apply Algorithm~\ref{newton} to estimate the $\alpha$ and $\beta$
parameters for every account using all data. The final influence measure is constructed by 
computing the influence measure vector $\hat{\Xi}$ over different time intervals 
to study how influence evolved; i.e. $\hat{\Xi}$ was computed by using the average $M_i(\mathcal{T}_m)$, where $\mathcal{T}_m$ denotes the $m-$th time interval
of interest.

\begin{figure}[!t]
\includegraphics[width=0.33\columnwidth,trim=2.5cm 2cm 1.8cm 1.2cm, clip=true]{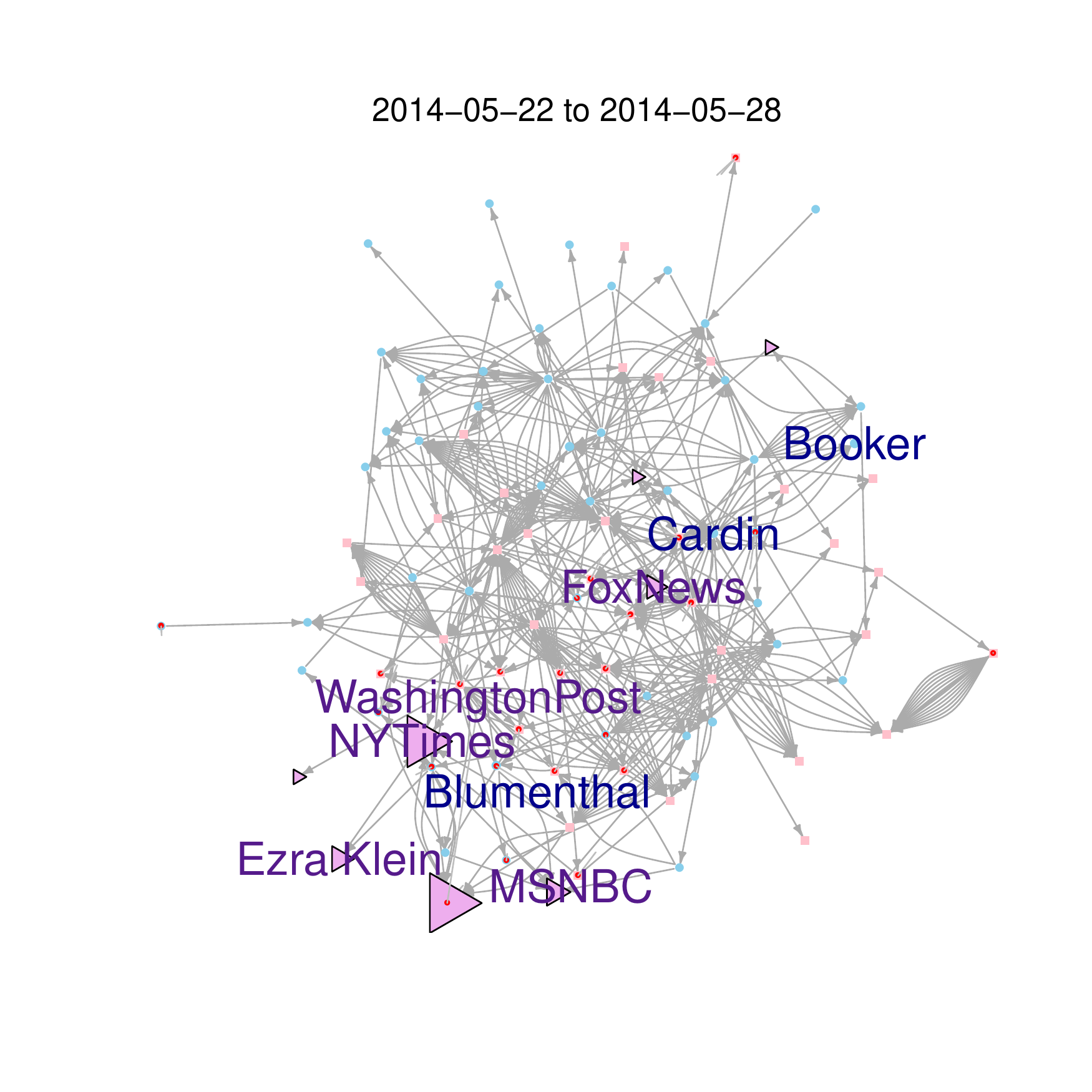}
\includegraphics[width=0.33\columnwidth,trim=2.5cm 1.8cm 1.8cm 1.2cm, clip=true]{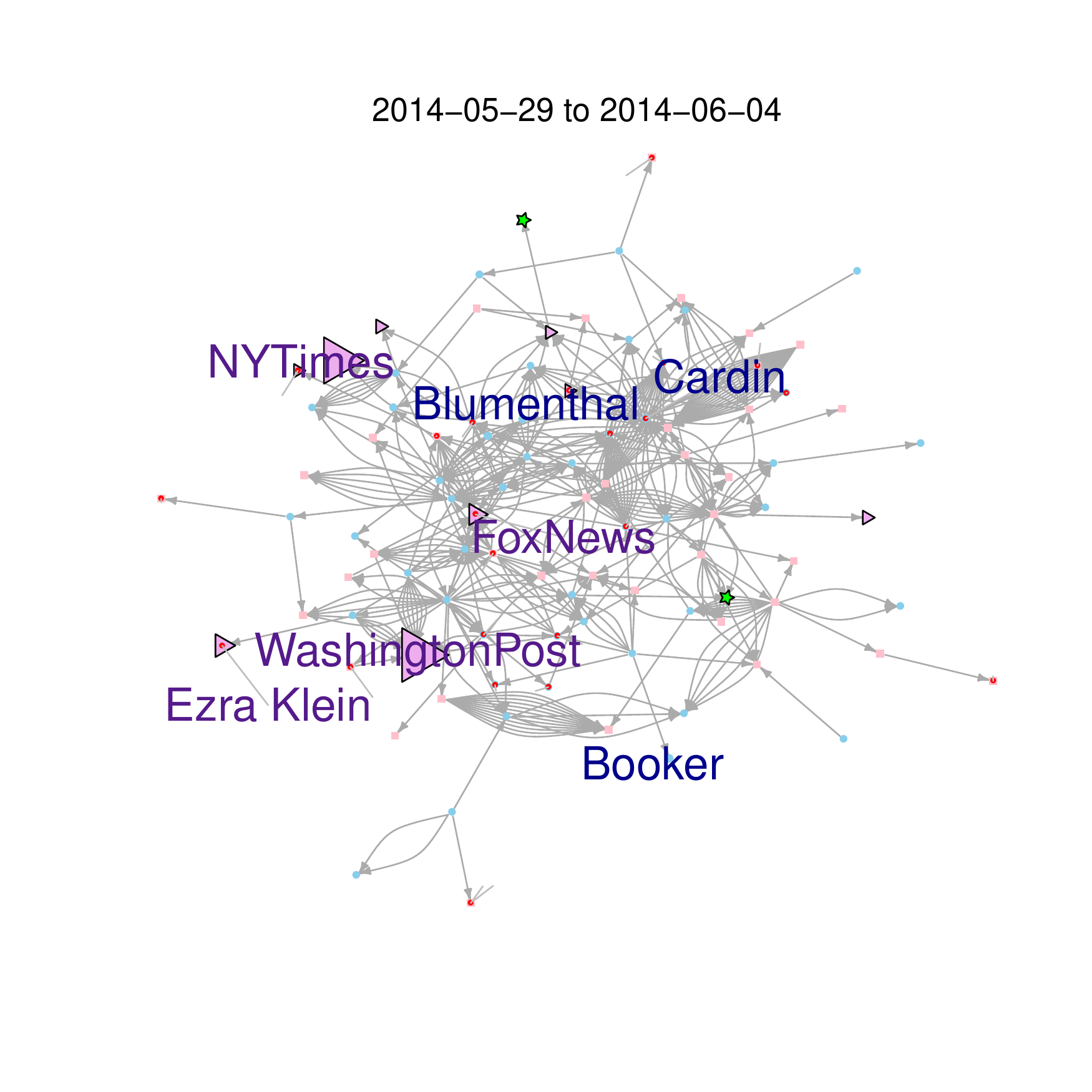}
\includegraphics[width=0.33\columnwidth,trim=2.5cm 1.8cm 1.8cm 1.2cm, clip=true]{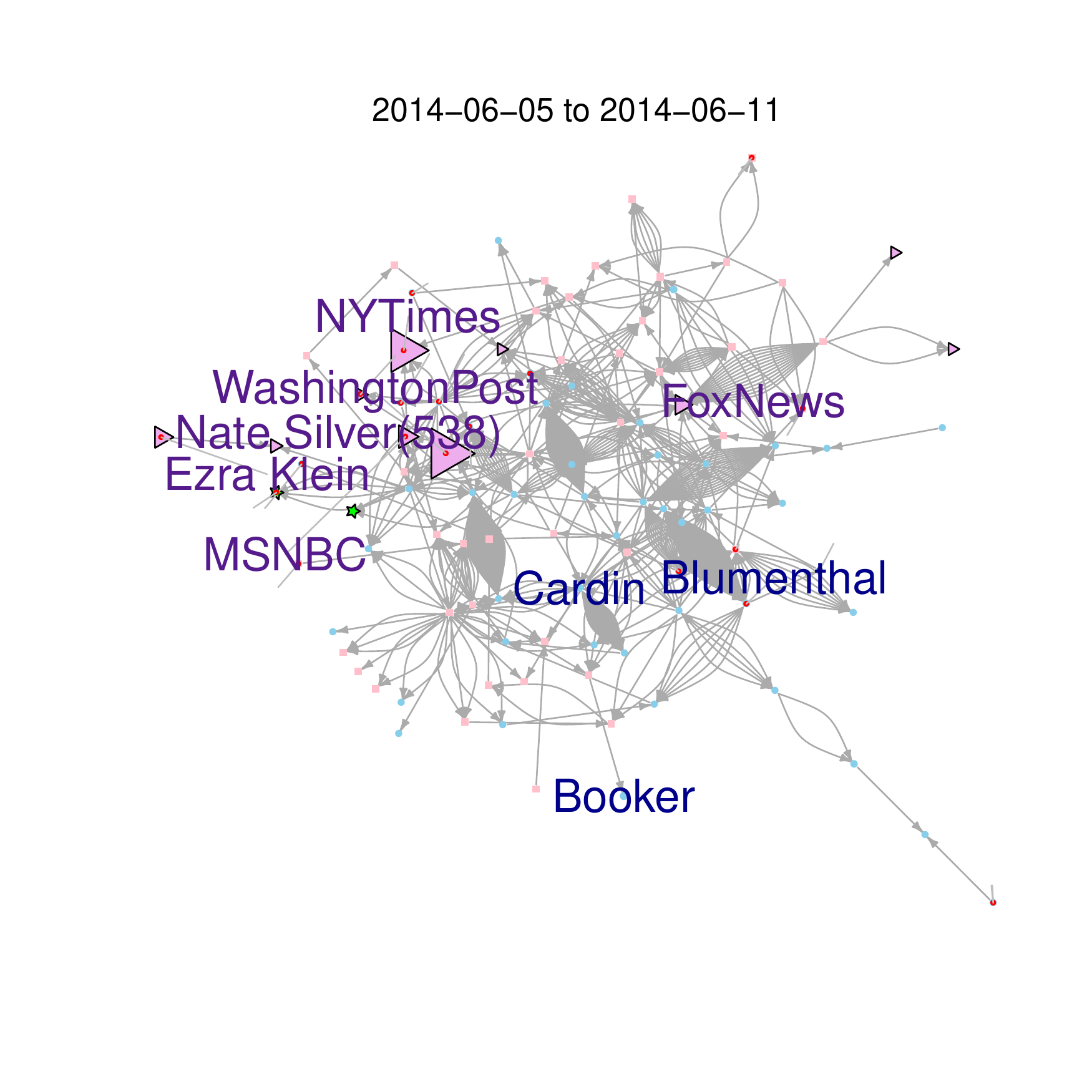}
\includegraphics[width=0.33\columnwidth,trim=2.5cm 2cm 1.8cm 1.2cm, clip=true]{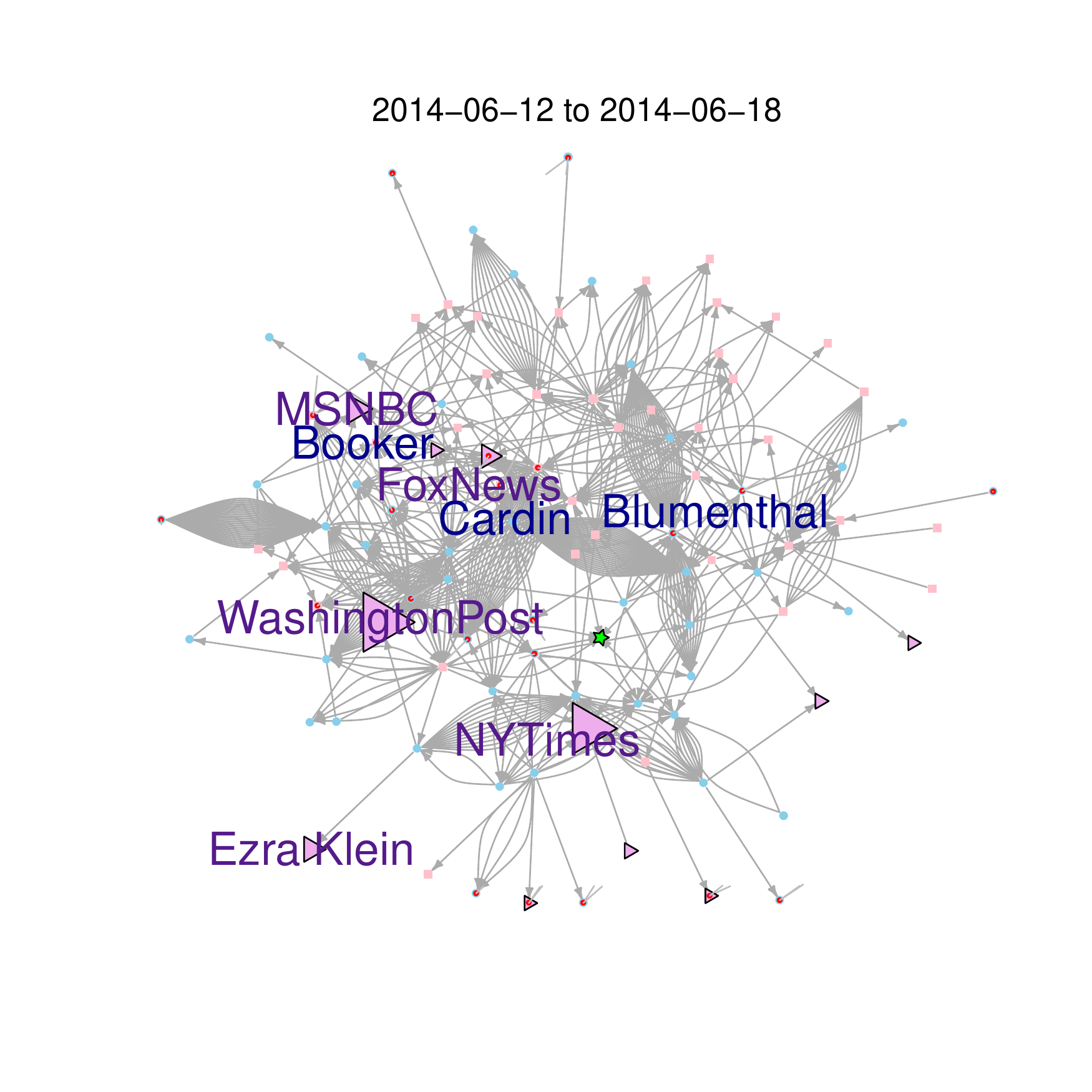}
\includegraphics[width=0.33\columnwidth,trim=2.5cm 2cm 1.8cm 1.2cm, clip=true]{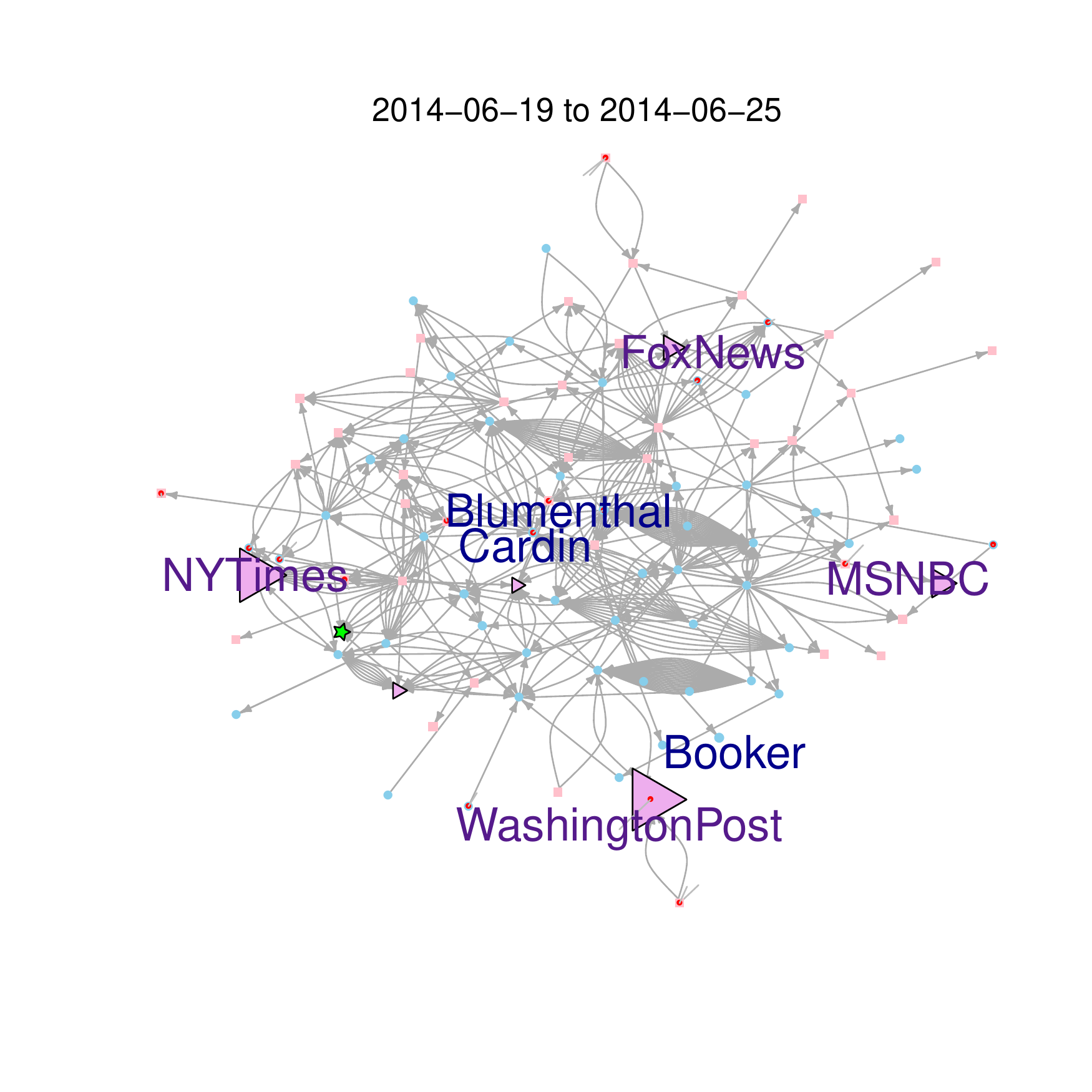}
\includegraphics[width=0.33\columnwidth,trim=2.5cm 1.8cm 1.8cm 1.2cm, clip=true]{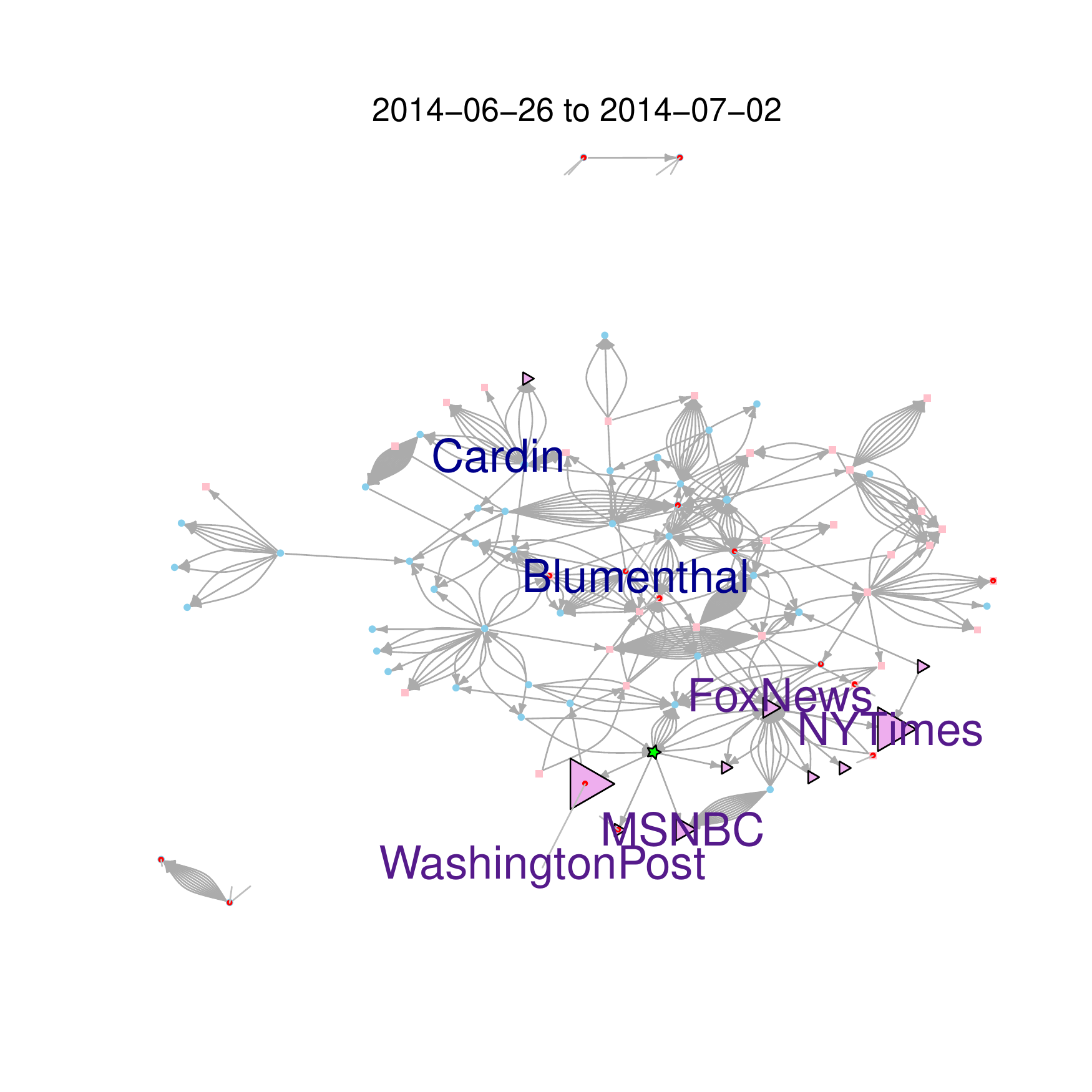}
\caption{Weekly Twitter retweet and mention network drawings for the 2014 summer. Top ten most influential accounts are labeled and node sizes are proportional to the estimated influence under the proposed model. The nodes (Twitter accounts) contain democratic senators (blue circles), republican senators (red squares), media (purple triangles), and government agencies (green stars).}
\label{fig:network_drawings_summer2014}
\end{figure}

The first time interval $\mathcal{T}_1$ we investigate is May 15, 2014 - July 3, 2014, which captures the most active period in our data and also represents a period when rate limiting is not a concern, i.e.,
 the data for even high volume users extends this far. During this time several major events occurred worldwide, including the soccer World Cup, 
debate on immigration reform, and the Islamic State in Iraq and the Levant (also known as the ISIS or ISIL) began 
an offensive in northern Iraq. Table~\ref{table:topten_summer2014} shows the 
top ten most influential accounts under the proposed method and PageRank 
\citep{page1999pagerank} calculated from the followers network. 
Both methods estimate that the Financial Times is the most influential Twitter account, and in general find that 
the media has an enormous influence that facilitates online conversation between politicians. We see from 
Figure~\ref{fig:network_drawings_summer2014} that these top accounts were actively retweeted and mentioned throughout this period.

\begin{table}
\centering
\caption{Top ten rankings according to the proposed model and PageRank from May 15, 2014 - July 3, 2014.}
\label{table:topten_summer2014}
\begin{tabular}{|r|l|l|}\hline
Rank & Proposed Measure & PageRank  \\\hline\hline
1 & Financial Times & Financial Times\\
2 & Washington Post & U.S. Army\\
3 & NYTimes & CNN\\
4 & MSNBC & Barack Obama\\
5 & Ezra Klein & CSPAN\\
6 & Fox News & New York Times\\
7 & Cory Booker & Washington Post\\
8 & Ben Cardin & Cory Booker \\
9 & Nate Silver (538) & MSNBC\\
10 & Richard Blumenthal & Wall Street Journal\\\hline
\end{tabular}
\end{table}

\begin{table}
\centering
\caption{Estimated R-squared values for different regression models, where the proposed measure and/or PageRank is included in the set of 
independent variables and the influence is computed for the entire data sample. We consistently find that the proposed measure is a better indicator of legislative importance.}
\label{table:R_squares}
\begin{tabular}{|l|c|c|r|}\hline
Response & Proposed Measure & PageRank & $R^{2}$ \\\hline\hline
 & $\check$ &  & 0.311 \\\cline{2-3}
leadership & & $\check$ & 0.276 \\\cline{2-3}
 & $\check$ & $\check$ & 0.311 \\\cline{2-3}\hline
 & $\check$ & & 0.114 \\\cline{2-3}
$log(\frac{\text{leadership}}{1 - \text{leadership}})$ & & $\check$ & 0.098 \\\cline{2-3}
 & $\check$ & $\check$ & 0.114 \\\cline{2-3}\hline
\end{tabular}
\end{table}

Next, we estimate the regression model in Equation~\ref{eqn:realdataRegn}. 
We note that Senators Baucus, Kerry, Cowan, Lautenberg, and Chiesa are scored by govtrack.us, but are not in our analysis. Max Baucus and John Kerry are left out, because they vacated their Senate seats to become, respectively, Ambassador to China and U.S. Secretary of State. 
Mo Cowan succeeded Kerry and was senator from February 1, 2013 to July 16, 2013 until a special election could be held. Cowan chose not to run in the election. Likewise, due to the death of Senator Frank Lautenberg, Jeffrey Chiesa was appointed by Governor Chris Christie to be the
junior senator from New Jersey from June 6, 2013 to October 31, 2013. He declined to run in the special election and thus, is also not included in the analysis. 

Since the leadership score provided by GovTrack are between $0$ and $1$, we estimate two models. One model uses the raw leadership scores, and 
another uses $log(\frac{\text{leadership}}{1 - \text{leadership}})$ for the response variable. In both cases, as shown in Table~\ref{table:R_squares},
we consistently find that the proposed influence measure explains more variation in leadership and when both the proposed and PageRank influence
 measures are included as independent variables, PageRank does not provide additional explanatory power. Tables~\ref{table:regression} and \ref{table:transformed_regression} 
show a significant positive coefficient for the proposed influence measure, meaning that senators who are more influential in Twitter by successfully steering conversation of their 
colleagues onto particular topics, tend to be more influential in real life in passing legislation. 
These results are consistent across different time intervals. For instance, in the Appendix we present similar results, where influence is calculated from January 1, 2013 to March 1, 2013 
corresponding to sequestration and also from November 1, 2012 to January 31, 2013 corresponding to the president's reelection and subsequent inauguration.

\begin{table}
\centering
\caption{Regression estimates, where the response variable is the raw leadership scores from GovTrack.us and influence is computed for the entire data sample. $R^{2}=0.311$; $F=8.228$ on $5$ and $92$ DF (p-value: $0.000$)}
\label{table:regression}
\begin{tabular}{|l|r|r|r|r|}\hline
Variable & Estimate & Std. Error & $t$ value & $P(>|t|)$ \\\hline\hline
Intercept & -0.086 & 0.232 & -0.368 & 0.714\\
Proposed Influence & 0.062 & 0.028 & 2.241 & 0.027\\
Republican & -0.154 & 0.039 & -3.945 & 0.000\\
Age & 0.002 & 0.003 & 0.923 & 0.359\\
Years in Senate & 0.007 & 0.003 & 2.518 & 0.014\\
Male & 0.020 & 0.050 & 0.395 & 0.694\\\hline
\end{tabular}
\end{table}

\begin{table}
\centering
\caption{Regression estimates, where the response variable is $log(\frac{\text{leadership}}{1 - \text{leadership}})$, where leaderhip is from GovTrack.us and influence is computed for the entire data sample. $R^{2}=0.114$; $F=2.334$ on $5$ and $92$ DF (p-value: $0.048$)}
\label{table:transformed_regression}
\begin{tabular}{|l|r|r|r|r|}\hline
Variable & Estimate & Std. Error & $t$ value & $P(>|t|)$ \\\hline\hline
Intercept & -3.590 & 2.604 & -1.379 & 0.171\\
Proposed Influence & 0.437 & 0.308 & 1.416 & 0.160\\
Republican & -1.112 & 0.438 & -2.538 & 0.013\\
Age & 0.009 & 0.029 & 0.323 & 0.747\\
Years in Senate & 0.034 & 0.032 & 1.063 & 0.290\\
Male & 0.470 & 0.563 & 0.834 & 0.407\\\hline
\end{tabular}
\end{table}

\section{Discussion}\label{discussion}

The goal in this paper was to characterize the influence of users 
in a large scale social media platform when given information about the 
detailed actions users take on it.  Our comprehensive analysis of the US Senators and 
related accounts demonstrated
that conversations, and in particular the rate of directed activity,
between accounts are correlated with their real-world position and influence.  We
expect similar conclusions to hold broadly for other types of directed
interaction data when the nodes form a clearly defined ecosystem or closely knit
social group/community.

The modeling and statistical inference issues, associated with these large scale data
are different from those in the related 
literature on network community detection  \citep{Kolaczyk, Fienberg, Murphy}, where the goal is to identify 
relatively dense groups of nodes (users), even though the underlying data  
(observed adjacency matrices) are the same. 
Relative to other recent work on  modeling directed networks, as in
\cite{perry2013point}, our study has important modeling differences
motivated by the online social media platform domain.  For instance,
our approach incorporates the fundamental differences between actions
like retweeting, mentioning, and posting. As a consequence, our final
influence measure, which sums all possible influences from the social
network, is able to outperform traditional topology driven approaches
like PageRank \citep{page1999pagerank}. Perhaps most importantly, given 
the massive volumes of data generated by platforms like Twitter, we 
presented a fast estimation algorithm and established 
statistical properties for the model estimates and those of the final
influence measure. 

These results suggest that the proposed model can be a relatively
straightforward technique to identifying influential individuals
within Twitter ecosystems, and that it can complement the significantly
more involved text mining based content analysis of the raw messages
for related tasks \citep{taddy}.

\makeatletter   
 \renewcommand{\@seccntformat}[1]{APPENDIX~{\csname the#1\endcsname}.\hspace*{1em}}
\makeatother

\section{Proofs} \label{appendix}

\subsection{Expressions for the gradient vector and Hessian matrix of the $LL$ function}

Some rather straightforward algebra yields the following expressions for the elements of the gradient vector $G\equiv\nabla_\Omega LL$:
\begin{equation}
  \begin{aligned}
\frac{\partial LL}{\partial \alpha_i}=\sum_{1\leq l\leq \Gamma} &\left\{ \sum_{1\leq j\leq n, j\neq i}\sum_{1\leq k \leq n_j}L_{ij} \log(M_i(T_{j,k}^l,l)+1)\right.\\
&\left.-\sum_{1\leq j\leq n}\sum_{1\leq k \leq n_j}\frac{\sum_{v\neq i}L_{iv} \log(M_i(T_{j,k}^l,l)+1)  \exp\left(\sum_{u\neq v}L_{uv}(\alpha_u+\beta_v) \log(M_u(T_{j,k}^l,l)+1) \right) } {\sum_{1\leq v\leq n} \exp\left(\sum_{u\neq v}L_{uv}(\alpha_u+\beta_v) \log(M_u(T_{j,k}^l,l)+1) \right)} \right\} 
\label{PA}
  \end{aligned}
\end{equation}
for $2\leq i\leq n$, and
\begin{equation}
  \begin{aligned}
\frac{\partial LL}{\partial \beta_j}=\sum_{1\leq l\leq \Gamma} &\left\{ \sum_{1\leq k \leq n_j}\sum_{i\neq j}L_{ij} M_i(T_{j,k}^l,l)\right.\\
&\left.-\sum_{1\leq s\leq n}\sum_{1\leq k \leq n_s}\frac{\left(\sum_{u\neq  j}L_{uj} \log(M_u(T_{s,k}^l,l)+1)\right)  \exp\left(\sum_{u\neq j}L_{uj}(\alpha_u+\beta_j) \log(M_u(T_{s,k}^l,l)+1) \right) } {\sum_{1\leq v\leq n} \exp\left(\sum_{u\neq v}L_{uv}(\alpha_u+\beta_v) \log(M_u(T_{s,k}^l,l)+1) \right)}
\right\} 
\label{PB}
  \end{aligned}
\end{equation}
for $1\leq j\leq n$. 

Next, we obtain the necessary expressions for the Hessian matrix $H(LL)$. We start by computing the sub-matrix of $H$ that includes the second partial derivatives
of $LL$ with respect to the $\alpha$ parameters. We get
\begin{equation}
  \begin{aligned}
\frac{\partial^2  LL}{\partial \alpha_i^2}=\sum_{1\leq l\leq \Gamma} &\left\{ -\sum_{1\leq j\leq n}\sum_{1\leq k \leq n_j}\frac{\sum_{v\neq i}L_{iv} \log(M_i^2(T_{j,k}^l,l)+1)  \exp\left(\sum_{u\neq v}L_{uv}(T_{j,k}^l)(\alpha_u+\beta_v) \log(M_u(T_{j,k}^l,l)+1) \right) } {\sum_{1\leq v\leq n} \exp\left(\sum_{u\neq v}L_{uv}(\alpha_u+\beta_v) \log(M_u(T_{j,k}^l,l)+1) \right)}\right.\\
&\left.+\sum_{1\leq j\leq n}\sum_{1\leq k \leq n_j}\frac{\left[\sum_{v\neq i}L_{iv} \log(M_i(T_{j,k}^l,l)+1)  \exp\left(\sum_{u\neq v}L_{uv}(\alpha_u+\beta_v) \log(M_u(T_{j,k}^l,l)+1) \right)\right]^2 } {\left[\sum_{1\leq v\leq n} \exp\left(\sum_{u\neq v}L_{uv}(\alpha_u+\beta_v) \log(M_u(T_{j,k}^l,l)+1) \right)\right]^2}\right\} 
\label{PAE}
  \end{aligned}
\end{equation}
When $i\neq q$, we similarly have
\begin{equation}
  \begin{aligned}
\frac{\partial^2  LL}{\partial \alpha_i\partial \alpha_q}=\sum_{1\leq l\leq \Gamma} &\left\{- \sum_{1\leq j\leq n}\sum_{1\leq k \leq n_j}\frac{\sum_{v\neq i,q}L_{iv} \log(M_i(T_{j,k}^l,l)+1)L_{qv} \log(M_q(T_{j,k}^l,l)+1)   } {\sum_{1\leq v\leq n} \exp\left(\sum_{u\neq v}L_{uv}(\alpha_u+\beta_v) \log(M_u(T_{j,k}^l,l)+1) \right)}\right.\\
& \cdot \exp\left(\sum_{u\neq v}L_{uv}(\alpha_u+\beta_v) \log(M_u(T_{j,k}^l,l)+1) \right)\\
&+\sum_{1\leq j\leq n}\sum_{1\leq k \leq n_j}\frac{\sum_{v\neq i}L_{iv} \log(M_i(T_{j,k}^l,l)+1)  \exp\left(\sum_{u\neq v}L_{uv}(\alpha_u+\beta_v) \log(M_u(T_{j,k}^l,l)+1) \right)} {\left[\sum_{1\leq v\leq n} \exp\left(\sum_{u\neq v}L_{uv}(\alpha_u+\beta_v) \log(M_u(T_{j,k}^l,l)+1) \right)\right]^2}\\
&\cdot \left[\sum_{v\neq q}L_{qv} \log(M_q(T_{j,k}^l,l)+1)  \exp\left(\sum_{u\neq v}L_{uv}(\alpha_u+\beta_v) \log(M_u(T_{j,k}^l,l)+1) \right)\right]\\
\label{PAD}
  \end{aligned}
\end{equation}
Next, we obtain the sub-matrix of $H$ that includes the second partial derivatives
of $LL$ with respect to the $\beta$ parameters and get
\begin{equation}
  \begin{aligned}
\frac{\partial^2 LL}{\partial \beta_j^2}=\sum_{1\leq l\leq \Gamma} &\left\{\sum_{1\leq s\leq n}\sum_{1\leq k \leq n_s}\frac{\left[\left(\sum_{u\neq  j}L_{uj} \log(M_u(T_{s,k}^l,l)+1)\right)  \exp\left(\sum_{u\neq j}L_{uj}(\alpha_u+\beta_j) \log(M_u(T_{s,k}^l,l)+1) \right) \right]^2} {\left[\sum_{1\leq v\leq n} \exp\left(\sum_{u\neq v}L_{uv}(\alpha_u+\beta_v) \log(N_u(T_{s,k}^l,l)+1) \right)\right]^2} \right.\\
&\left.-\sum_{1\leq s\leq n}\sum_{1\leq k \leq n_s}\frac{\left(\sum_{u\neq  j}L_{uj} \log(M_u(T_{s,k}^l,l)+1)\right)^2  \exp\left(\sum_{u\neq j}L_{uj}(\alpha_u+\beta_j) \log(M_u(T_{s,k}^l,l)+1) \right) } {\sum_{1\leq v\leq n} \exp\left(\sum_{u\neq v}L_{uv}(\alpha_u+\beta_v) \log(M_u(T_{s,k}^l,l)+1) \right)}
\right\} 
\label{PBE}
  \end{aligned}
\end{equation}
for $1\leq j\leq n$. When $j\neq q$, we can similarly have
\begin{equation}
  \begin{aligned}
\frac{\partial^2 LL}{\partial \beta_j\partial\beta_q}=\sum_{1\leq l\leq \Gamma} &\left\{\sum_{1\leq s\leq n}\sum_{1\leq k \leq n_s}\frac{\left(\sum_{u\neq  j}L_{uj}\log(M_u(T_{s,k}^l,l)+1)\right)  \exp\left(\sum_{u\neq j}L_{uj}(\alpha_u+\beta_j) \log(M_u(T_{s,k}^l,l)+1) \right) } {\left[\sum_{1\leq v\leq n} \exp\left(\sum_{u\neq v}L_{uv}(\alpha_u+\beta_v) \log(M_u(T_{s,k}^l,l)+1) \right)\right]^2}\right.\\
&\left. \cdot  \left(\sum_{q\neq  j}L_{uq} \log(M_u(T_{s,k}^l,l)+1)\right)  \exp\left(\sum_{u\neq q}L_{uj}(\alpha_u+\beta_q) \log(M_u(T_{s,k}^l,l)+1) \right)\right\}
\label{PBD}
  \end{aligned}
\end{equation}

Finally, we provide expressions for the cross-partials
\begin{equation}
  \begin{aligned}
\frac{\partial^2 LL}{\partial \alpha_i \partial \beta_i}=\sum_{1\leq l\leq \Gamma} &\left\{ \sum_{1\leq s\leq n}\sum_{1\leq k \leq n_s}\frac{\left(\sum_{u\neq  i}L_{ui}\log(M_u(T_{s,k}^l,l)+1)\right) \exp\left(\sum_{u\neq i}L_{ui}(\alpha_u+\beta_i) \log(M_u(T_{s,k}^l,l)+1) \right) } {\left[\sum_{1\leq v\leq n} \exp\left(\sum_{u\neq v}L_{uv}(\alpha_u+\beta_v) \log(M_u(T_{s,k}^l,l)+1) \right)\right]^2}\right.\\
& \left. \cdot \sum_{v\neq i}L_{iv} \log(M_i(T_{s,k}^l,l)+1)  \exp\left(\sum_{u\neq v}L_{uv}(\alpha_u+\beta_v) \log(M_u(T_{s,k}^l,l)+1) \right) 
\right\} 
\label{PABE}
  \end{aligned}
\end{equation}
When $i\neq j$, 

\small
\begin{equation}
  \begin{aligned}
\frac{\partial^2 LL}{\partial \alpha_i \partial \beta_j}&=\sum_{1\leq l\leq \Gamma} \left\{ \sum_{1\leq s\leq n}\sum_{1\leq k \leq n_s}\frac{\left(\sum_{u\neq  j}L_{uj} \log(M_u(T_{s,k}^l,l)+1)\right) \exp\left(\sum_{u\neq j}L_{uj}(\alpha_u+\beta_j) \log(M_u(T_{s,k}^l,l)+1) \right) } {\left[\sum_{1\leq v\leq n} \exp\left(\sum_{u\neq v}L_{uv}(\alpha_u+\beta_v) \log(M_u(T_{s,k}^l,l)+1) \right)\right]^2}\right.\\
& \cdot \sum_{v\neq i}L_{iv}\log(M_i(T_{s,k}^l,l)+1)  \exp\left(\sum_{u\neq v}L_{uv}(\alpha_u+\beta_v) \log(M_u(T_{s,k}^l,l)+1) \right) \\
&\left.-\sum_{1\leq s\leq n}\sum_{1\leq k \leq n_s}\frac{\left(\sum_{u\neq  j}L_{uj}\log(M_u(T_{s,k}^l,l)+1)\right) L_{ij}\log(M_i(T_{s,k}^l,l)+1) \exp\left(\sum_{u\neq j}L_{uj}(\alpha_u+\beta_j) \log(M_u(T_{s,k}^l,l)+1) \right) } {\sum_{1\leq v\leq n} \exp\left(\sum_{u\neq v}L_{uv}(\alpha_u+\beta_v) \log(M_u(T_{s,k}^l,l)+1) \right)}
\right\} 
\label{PABD}
  \end{aligned}
\end{equation}

\subsection{Proof of Theorem 1}

 Note that since the baseline hazard rate $\lambda_{0,l}(t)$ is canceled out in the partial likelihood and the estimation of the parameter vector 
$\Omega$ will not depend on its value, in the rest of the proof, we can safelyl ignore it. Next, consider the process
\[X(\Omega',t)=\Gamma^{-1}(LL(\Omega', t)-LL(\Omega,t)),\]
where recall that $\Gamma$ denotes the number of topics under consideration. 
Then, given $\Omega$, it is straightforward to see that $\hat{\Omega}$ is the unique maximum point (with probability going to 1)  of
$X(\Omega, t_0)$ \citep{andersen1982cox} and $X(\Omega, t_0)$ is a concave function. To simplify notation,
we let $TN(t,l)=\sum_i N_i(t,l)$. Then, we can expand $X(\Omega',t)$ as:
\begin{equation*}
  \begin{aligned}
X(\Omega',t) &=\Gamma^{-1}\left\{\sum_l\int_0^t \sum_j \sum_{i, i\neq j}L_{ij}(\alpha_i'+\beta_j'-\alpha_i-\beta_j)\log(M_i(u,l)+1)dN_j(u,l) \right.\\
&\left.-\int_0^t \log \left\{\frac{\sum_j \exp\left(\sum_{i,i\neq j}L_{ij}(\alpha_i'+\beta_j')  \log(M_i(u,l)+1) \right)}{\sum_j \exp\left(\sum_{i, i\neq j}L_{ij}(\alpha_i+\beta_j)\log(M_i(u,l)+1)) \right)}\right\} dTN(u,l) \right\}
  \end{aligned}
\end{equation*}

Notice that by the definitions of the hazard rates, by defining
\begin{equation}
E_j(t,l)=N_j(t,l)-\int_0^t \exp\left(\sum_{i,i\neq j}L_{ij}(u)(\alpha_i+\beta_j) \log(M_i(u,l)+1)\right)du  \label{SMD}
\end{equation}
$j=1,\ldots, n$,  we can easily establish that $E_j(t,l)$s are  local martingales on the time interval $[0,t_0]$. 
Then, in $X(\Omega',t)$,  we replace  $d N_j(t,l)$  with the hazard rates of $N_j(t,l)$, which is
\begin{equation*}
  \begin{aligned}
&\exp\left(\sum_{i\neq j}L_{ij}(t)(\alpha_i+\beta_j) \log(M_i(t,l)+1) \right)\\
  \end{aligned}
\end{equation*}

Then, we can get
\small
\begin{equation}
  \begin{aligned}
A(\Omega',t) &=\Gamma^{-1}\left\{\sum_l\int_0^t \sum_j \sum_{i, i\neq j}L_{ij}(\alpha_i'+\beta_j'-\alpha_i-\beta_j)\log(M_i(u,l)+1) \exp\left(\sum_{i\neq j}L_{ij}(t)(\alpha_i+\beta_j) \log(M_i(t,l)+1) \right)du\right.\\
&\left. -\int_0^t \log \left\{\frac{\sum_j \exp\left(\sum_{i,i\neq j}L_{ij}(\alpha_i'+\beta_j')  \log(M_i(u,l)+1) \right)}{\sum_j \exp\left(\sum_{i, i\neq j}L_{ij}(\alpha_i+\beta_j)\log(M_i(u,l)+1) \right)}\right\} \left(\sum_{j}\exp\left(\sum_{i\neq j}L_{ij}(\alpha_i+\beta_j) \log(M_i(t,l)+1) \right)\right)du\right\}
\label{MD}
  \end{aligned}
\end{equation}
\normalsize

Some algebra shows that $X(\Omega', t)-A(\Omega', \cdot)$ can be written as a finite sum of the square integrable  local martingales
 $E_j(t,l)$ in (\ref{SMD}). Then, the lastter is also a local square integrable martingale. By Theorem 2.4.3 in \citep{fleming}, we have 
\[<X(\Omega', t)-A(\Omega', t), X(\Omega', t)-A(\Omega', t)>=B(\Omega', t),\]
 where 
\begin{equation}
B(\Omega', t)=\Gamma^{-2}\sum_l \sum_{j} \ \int_0^t S_j(u,l)^2 \lambda_0 \exp\left(\sum_{i, i\neq j}L_{ij}(\alpha_i+\beta_j) \log(M_i(u,l)+1) \right)du,
\label{FV}
\end{equation}
where in  (\ref{FV}) above, $S_j(u,l)$ is given by
\begin{equation*}
S_j(u,l)=\sum_{i, i\neq j}L_{ij}(\alpha_i'+\beta_j'-\alpha_i-\beta_j)\log(M_i(u,l)+1)- \log \left\{\frac{\sum_j \exp\left(\sum_{i,i\neq j}L_{ij}(\alpha_i'+\beta_j')\log(M_i(u,l)+1)\right)}{\sum_j \exp\left(\sum_{i, i\neq j}L_{ij}(\alpha_i+\beta_j) \log(M_i(u,l)+1) \right)}\right\}
\end{equation*}

Since $\Gamma \cdot B(\Omega', \cdot)$ converges in probability to some finite quantity (by the  law of large numbers, depending on $\Omega$), $X(\Omega', t)$ should converge to the same limit as $A(\Omega',t)$ for each $\Omega'$ (if $A(\Omega',t)$ converges). Note that by using
the notation employed in Condition (A) and (B) in Theorem 1, $A(\Omega',t)$ can be simplified to
\begin{equation*}
  \begin{aligned}
A(\Omega', t)=& \int_0^t  \left[\sum_j (\Phi_j'-\Phi_j)'E_j^{(1)}(u,\Omega)-\log\left\{\frac{\sum_j E_j(u,\Omega')}{\sum_j E_j(u, \Omega)}\right\}\sum_j E_j(u,\Omega)\right]du \\
  \end{aligned}
\end{equation*}

It follows that by conditions (A), (B) and (C), for each $\Omega'$, as $\Gamma\to \infty$, 
\[A(\Omega', t_0)\to_P P(\Omega', t_0),\]
where
\[P(\Omega', t_0)=\int_0^{t_0}  \left[\sum_j (\Phi_j'-\Phi_j)'e_j^{(1)}(u, \Omega)-\log\left\{\frac{\sum_j e_j(u,\Omega')}{\sum_j e_j(u, \Omega)}\right\}\sum_j e_j(u,\Omega)\right]du \]

Now, we want to establish that $P(\Omega', t_0)$ remains convex. We evaluate the first and second derivative of $P_1(\Omega', t_0)$ to show its convexity. By Condition (C), we can compute the first derivatives as
\[\frac{\partial P(\Omega', t_0)}{\partial \beta_j'}=\int_0^{t_0}  \left[e_j^{(1)}(u, \Omega)-e_j^{(1)}(u,\Omega')\frac{\sum_j e_j(u,\Omega)}{\sum_j e_j(u, \Omega')} \right]du \]
and
\[\frac{\partial P(\Omega', t_0)}{\partial \alpha_i'}=\int_0^{t_0}  \left[\sum_j I_i' e_j^{(1)}(u, \Omega)-(\alpha_1',\ldots, \alpha_n')'\sum_j I_i' e_j^{(1)}(u,\Omega') \frac{\sum_j e_j(u,\Omega)}{\sum_j e_j(u, \Omega')} \right]du \]
where $I_i$ is a $n$-dimensional vector with all zeros expect one on the $i$-th entry. 

Note that the above parital derivatives are all zero at $\Omega'=\Omega$. Further, the second derivatives are
\[\int_0^{t_0}  \left[e_j^{(2)}(u, \Omega)+e_j^{(1)}(u, \Omega)^{\otimes 2}\frac{\sum_j e_j(u,\Omega)}{\sum_j e_j(u, \Omega')}\right]du,\]
which is a positive semidefinite matrix for any $\Omega'$ and positive definite at $\Omega$. Thus, $X(\Omega', t_0)$ converges in probability to a convex function of $\Omega'$ with a unique maximum at $\Omega$. Since $\hat{\Omega}$ maximizes the concave function $X(\Omega', t_0)$, it follows by 
a standard result in convex analysis \citep{rockafellar} that $\hat{\Omega}\to_P \Omega$. This completes the proof of the Theorem.

\section{Additional Senator Results}

Table~\ref{table:topten_all} shows the 
top ten most influential accounts under the proposed method for different time periods. 
We see consistent results with the findings from summer 2014. Important newspapers like 
the Financial Times and Washington Post still appear in the top ten when utilizing the full 
data. Other prominent accounts include senators that have leadership positions, like Harry 
Reid (Senate Majority Leader) and several others with high profile committee chairmanships 
or ranking appointments. 

\begin{table}
\centering
\caption{Top ten rankings under the proposed model for different time intervals.}
\label{table:topten_all}
\begin{tabular}{|r|l|l|l|}\hline
Rank & Sequestration & 2014 Inauguration & Entire Data  \\\hline\hline
1 & Leahy & Leahy & Financial Times\\
2 & Grassley & Grassley& Grassley \\
3 & Mikulski & Begich & Leahy\\
4 & Begich & Mikulski & Cruz\\
5 & Shaheen & Johanns & Washington Post\\
6 & McCaskill & Reid& Reid \\
7 & Reid & McCaskill& Begich \\
8 & Blunt & Graham & Mikulski \\
9 & Graham & Shaheen & Ezra Klein\\
10 & Collins & Hagan & Schatz\\\hline
\end{tabular}
\end{table}

Tables~\ref{table:regression_sequestration} and \ref{table:transformed_regression_sequestration} show regression results for the sequestration period, and Tables~\ref{table:regression_inauguration} and \ref{table:transformed_regression_inauguration} show regression results for the inauguration period. The results are consistent with the results presented in the main text. Regressing directly on the leadership scores shows a strongly significant and positive coefficient for the proposed influence measure. The regressions with transformed leadership scores show effects are moderately significant.

\begin{table}
\centering
\caption{Regression estimates, where the response variable is the raw leadership scores from GovTrack.us and influence is computed from January 1, 2013 to March 1, 2013. $R^{2}=0.327$; $F=8.839$ on $5$ and $92$ DF (p-value: $0.000$)}
\label{table:regression_sequestration}
\begin{tabular}{|l|r|r|r|r|}\hline
Variable & Estimate & Std. Error & $t$ value & $P(>|t|)$ \\\hline\hline
Intercept & -0.153 & 0.228 & -0.669 & 0.505\\
Proposed Influence & 0.074 & 0.028 & 2.689 & 0.009\\
Republican & -0.153 & 0.039 & -3.960 & 0.000\\
Age & 0.002 & 0.003 & 0.833 & 0.407\\
Years in Senate & 0.007 & 0.003 & 2.532 & 0.013\\
Male & 0.020 & 0.050 & 0.397 & 0.692\\\hline
\end{tabular}
\end{table}

\begin{table}
\centering
\caption{Regression estimates, where the response variable is $log(\frac{\text{leadership}}{1 - \text{leadership}})$, where leaderhip is from GovTrack.us and influence is computed from January 1, 2013 to March 1, 2013. $R^{2}=0.119$; $F=2.466$ on $5$ and $92$ DF (p-value: $0.038$)}
\label{table:transformed_regression_sequestration}
\begin{tabular}{|l|r|r|r|r|}\hline
Variable & Estimate & Std. Error & $t$ value & $P(>|t|)$ \\\hline\hline
Intercept & -3.925 & 2.574 & -1.525 & 0.131\\
Proposed Influence & 0.504 & 0.312 & 1.614 & 0.110\\
Republican & -1.103 & 0.437 & -2.526 & 0.013\\
Age & 0.008 & 0.029 & 0.267 & 0.790\\
Years in Senate & 0.033 & 0.031 & 1.062 & 0.291\\
Male & 0.465 & 0.561 & 0.830 & 0.409\\\hline
\end{tabular}
\end{table}

\begin{table}
\centering
\caption{Regression estimates, where the response variable is the raw leadership scores from GovTrack.us and influence is computed from November 1, 2012 to January 31, 2013. $R^{2}=0.328$; $F=8.839$ on $5$ and $92$ DF (p-value: $0.000$)}
\label{table:regression_inauguration}
\begin{tabular}{|l|r|r|r|r|}\hline
Variable & Estimate & Std. Error & $t$ value & $P(>|t|)$ \\\hline\hline
Intercept & -0.132 & 0.220 & -0.597 & 0.552\\
Proposed Influence & 0.072 & 0.026 & 2.726 & 0.008\\
Republican & -0.154 & 0.039 & -3.978 & 0.000\\
Age & 0.002 & 0.003 & 0.797 & 0.427\\
Years in Senate & 0.007 & 0.003 & 2.616 & 0.010\\
Male & 0.020 & 0.050 & 0.395 & 0.693\\\hline
\end{tabular}
\end{table}

\begin{table}
\centering
\caption{Regression estimates, where the response variable is $log(\frac{\text{leadership}}{1 - \text{leadership}})$, where leaderhip is from GovTrack.us and influence is computed from November 1, 2012 to January 31, 2013. $R^{2}=0.117$; $F=2.402$ on $5$ and $92$ DF (p-value: $0.043$)}
\label{table:transformed_regression_inauguration}
\begin{tabular}{|l|r|r|r|r|}\hline
Variable & Estimate & Std. Error & $t$ value & $P(>|t|)$ \\\hline\hline
Intercept & -3.578 & 2.495 & -1.434 & 0.155\\
Proposed Influence & 0.452 & 0.297 & 1.521 & 0.132\\
Republican & -1.105 & 0.437 & -2.527 & 0.013\\
Age & 0.007 & 0.029 & 0.255 & 0.800\\
Years in Senate & 0.035 & 0.031 & 1.113 & 0.269\\
Male & 0.460 & 0.561 & 0.819 & 0.415\\\hline
\end{tabular}
\end{table}

\bibliographystyle{ECA_jasa}
\bibliography{biblio}

\end{document}